\documentclass[preprint,aps,pre,letterpaper,floatfix,showpacs]{revtex4}
\usepackage{graphicx,amsmath}
\usepackage{graphicx}
\usepackage{dcolumn}
\usepackage{bm}
\usepackage{url}

\def\I.#1{\it #1}
\def\B.#1{{\bbox#1}}
\def\C.#1{{\cal #1}}

\newcommand\eq{\begin{equation}}
\newcommand\en{\end{equation}}

\newcommand\tix{{\tilde{x}}}
\newcommand\tiy{{\tilde{y}}}

\begin{document}

\title{On the stabilization of ion sputtered surfaces}
\author{Benny Davidovitch$^{1,2}$, Michael J. Aziz$^1$, and Michael P. Brenner$^1$}
\affiliation{$^1$ Harvard School of Engineering and Applied Sciences,
Cambridge, MA 02138 \\
$^2$ Physics Department, University of Massachusetts, Amherst, MA
01002}
\date{\today}
\begin{abstract}
The classical theory of ion beam sputtering predicts the instability
of a flat surface to uniform ion irradiation at any incidence angle.
We relax the assumption of the classical theory that the average
surface erosion rate is determined by a Gaussian response function
representing the effect of the collision cascade and consider the
surface dynamics for other physically-motivated response functions.
We show that although instability of flat surfaces at any beam angle
results from all Gaussian and a wide class of non-Gaussian erosive
response functions, there exist classes of modifications to the
response that can have a dramatic effect. In contrast to the
classical theory, these types of response render the flat surface
linearly stable, while imperceptibly modifying the predicted sputter
yield vs. incidence angle. We discuss the possibility that such
corrections underlie recent reports of a ``window of stability'' of
ion-bombarded surfaces at a range of beam angles for certain ion and
surface types, and describe some characteristic aspects of pattern
evolution near the transition from unstable to stable dynamics. We
point out that careful analysis of the transition regime may provide
valuable tests for the consistency of any theory of pattern
formation on ion sputtered surfaces. 
\end{abstract}
\maketitle
\section{Introduction}
Uniform ion beam sputter erosion of a solid surface often causes a
spontaneously-arising topographic pattern in the surface topography
\cite{Navez62,Carter77,Lewis80,Bradley88,Malherbe94,Mayer94,Chason94a,
Cuerno95,Carter96,Makeev97,Facsko99,Judy99,Erlebacher99,
Erlebacher00,Facsko01,Flamm01,Habenicht01,Costantini01,Umbach01,Valbusa02,
Makeev02,Chason03,Facsko04,Castro05,Chan05,Cuenat05,Feix05,Rusponi05,
Brown05a,Brown05b,Ziberi05,Ziberi06a,Aziz06}, that can take the form
of a one-dimensional corrugation or a two-dimensional array of dots
with typical length scales of $10^{2 \pm 1}$ nm. Periodic
self-organized patterns with wavelength as small as 15 nm
\cite{Mayer94,Facsko01} have stimulated interest in this method as a
means of nanofabrication at sub-lithographic length scales
\cite{Cuenat05}. Because the characteristic scale of the patterns
can be three orders of magnitude larger than the characteristic
penetration depth of ions into a solid surface, the patterns result
from a nontrivial interplay between the sputter erosion on one hand
and surface relaxation mechanisms on the other hand.

The present understanding of sputter morphology evolution originates
in the Sigmund theory of sputtering \cite{Sigmund69}.  Sigmund
posited that the local erosion rate of the surface is proportional
to the local atom emission rate resulting from the atomic collision
cascade, and that the emission rate at a point on the surface is
proportional to the nuclear energy deposition density at that point
resulting from collision cascades from the ions impinging at all
points. Sigmund  subsequently \cite{Sigmund73} recognized the
destabilizing influence of the curvature-dependence of the sputter
yield (atoms out per incident ion) by modeling the nuclear energy
deposition density as taking the form of Gaussian ellipsoids beneath
the surface and showing that, as a consequence, concave regions of
the surface receive more energy and thereby erode more rapidly than
do convex regions \footnote{For crystallographic surfaces below the
thermodynamic roughening transition temperature, non-classical
surface diffusion can be a similarly destabilizing influence
\cite{Villain91}.}.

The origin of the characteristic length scale of the self-organized
patterns was identified by Bradley and Harper (BH) \cite{Bradley88}, who recognized
that Sigmund's destabilization mechanism is opposed by surface
diffusion, which operates so as to return the surface to flatness\footnote{For amorphous materials, ion-stimulated viscous flow can
be a similarly stabilizing influence \cite{Umbach01}.}. Expanding
Sigmund's Gaussian ellipsoid response in powers of derivatives of the surface height
$h(x,y,t)$ and superposing classical Mullins-Herring
\cite{Herring50,Mullins59} surface diffusion, BH derived a linear partial
differential equation (PDE) \cite{Bradley88} that describes the
evolution of the surface height on scales much larger than the
characteristic length scales of Sigmund's Gaussian response:
\begin{equation}
\frac{\partial h}{\partial t}  = - I + \{ S_x
\partial_{xx} + S_y \partial_{yy}  -
 B \nabla^4 \} h  \ , \label{eq0a}
\end{equation}
where $I(b)$ is the vertical erosion rate of a flat surface,
$S_x(b)$ and $S_y(b)$ are the curvature coefficients, $b$ is the
surface slope, and $B$ is a material parameter describing relaxation
and containing the surface diffusivity and the surface free energy.
The coefficients $I, S_{x}$, and $S_{y}$ are expressed in terms of
Sigmund's Gaussian and depend on $\theta=\tan^{-1}(b)$, the angle
between the beam direction, henceforth denoted as $-\hat{z}$, and
the local normal to the surface $\hat{n}$ ($0 \leq \theta \leq
\pi/2$). For nonzero $\theta$, we denote by $\hat{x}$ the axis
perpendicular to $\hat{z}$ in the $\hat{n}-\hat{z}$ plane.
Bradley-Harper's linear stability analysis yields unstable modes
whenever $S_x$ or $S_y$ is negative, whose characteristic length
scale arises from a balance between the destabilizing effect of the
second derivatives $\partial_{xx} ,\partial_{yy}$ and the
stabilizing effect of the surface diffusion term $\nabla^4$. The
behavior of $S_x(\theta)$ and $S_y(\theta)$ for characteristic
parameter values are shown in Fig. 1.
\begin{figure}
\begin{center}
\includegraphics[width=2.0in,clip=]{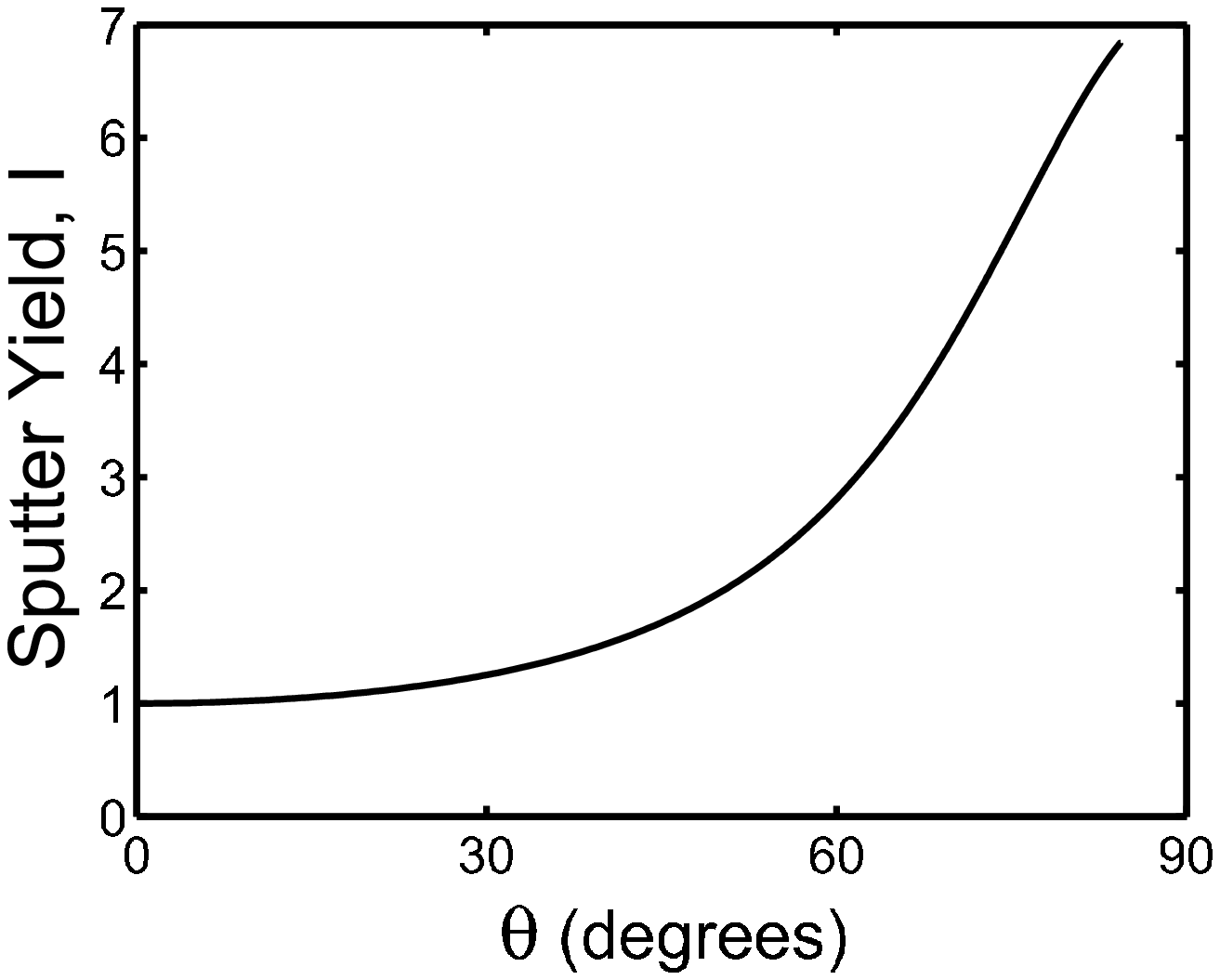}
\includegraphics[width=2.0in,clip=]{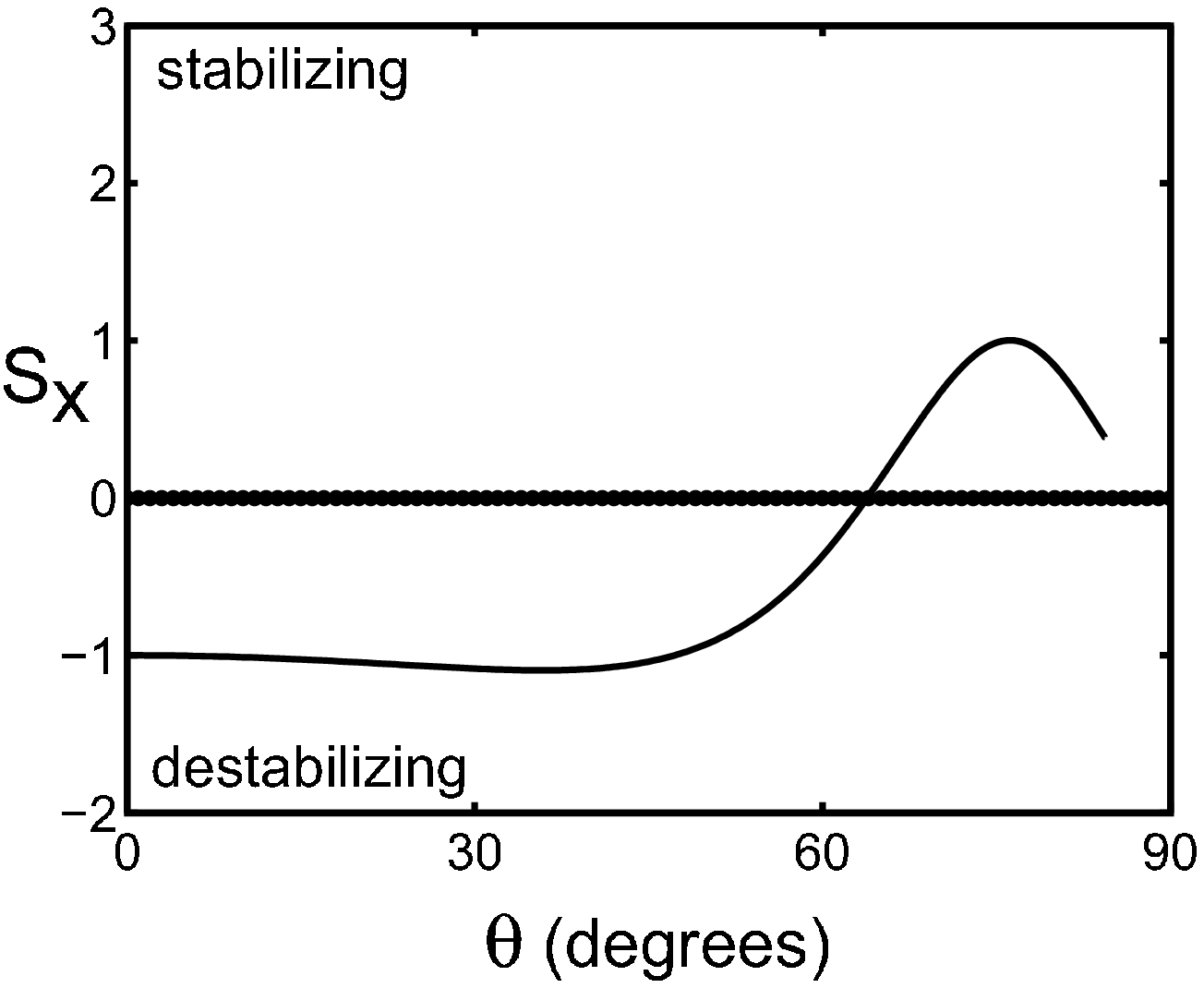}
\includegraphics[width=2.0in,clip=]{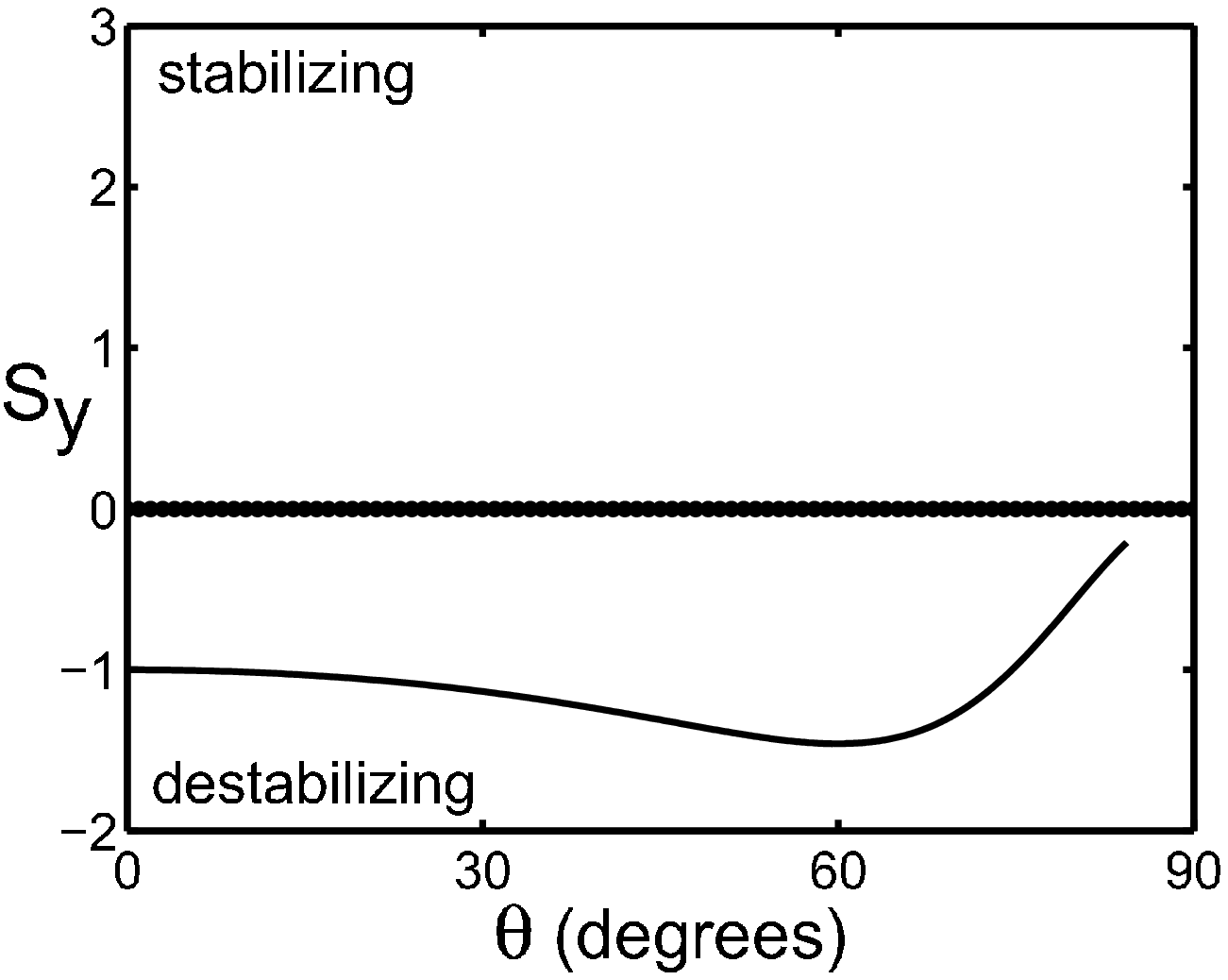}
\caption{\label{figBH} (a) Plot of sputter yield curve $I(\theta)$,
normalized by $I(0)$ (b,c) Plots of $S_x(\theta)$ and $S_y(\theta)$,
normalized by $|S_x(0 )|=|S_y(0)|$. The parameters used are: $a=1.5$
nm, $\sigma=0.9$ nm, $\mu=0.5$ nm.}
\end{center}
\end{figure}
The Bradley-Harper analysis gives rise to the following predictions:
(i) Below a crossover angle $\theta_{cross} $, $S_{x} < S_{y} <0$,
implying a faster growth rate for parallel mode (wave vector
parallel to projected ion beam direction along the surface) than for
perpendicular mode (wave vector along $\hat{y}$) surface modulations
\footnote{This conclusion depends on the relative values of
$a,\mu,\sigma$, but is generally correct for the values achieved in
practice.}; (ii) $S_{y} < 0$ for all $\theta$, implying instability
to perpendicular modes at \emph{all} incidence angles. For $\theta
>\theta_{cross}$ the perpendicular modes are the fastest to grow with
dominant wavelength $\sqrt{8\pi^2B/(-S_y)}$. The generalization of
the BH analysis to the nonlinear regime, which is required to
account for the observed saturation of ripple amplitude and the
emergence of more complicated patterns (e.g. hexagons, dots, pits)
was carried out by Cuerno and coworkers \cite{Makeev02,Cuerno95} who
expanded Sigmund's Gaussian ellipsoid model to higher order in
surface height derivatives, resulting in a Kuramoto-Sivashinsky type
equation \cite{Cross93} for the surface evolution.

There is growing evidence that although the Bradley-Harper
predictions explain some features of experiments (e.g. the
temperature dependence of the wavelength of the ripples
\cite{Erlebacher99}), there are also some glaring inconsistencies.
This is clearly demonstrated in, e.g., the recent work of Ziberi
{\it et al.} \cite{Ziberi05}, who found a ``window of stability''
for Si surfaces at room temperature bombarded by $\sim 1-2$ keV
noble gas ions at an intermediate range of angles $\theta_1 <\theta
<\theta_2$, where $\theta_1 \approx 30^o$, and $\theta_2 \approx
60^o$. Moreover, Ziberi {\it et al.} demonstrated that when
bombarded by some noble ions (${\rm Ne^{+}}$), a flat surface
remains stable at all angles. In addition to the experimental
inconsistencies with BH prediction (ii), there have also been recent
experiments \cite{Costantini01,Yamada01} and atomistic simulations
\cite{Bringa01,Naga07} that have measured the shape change of a
smooth solid surface in the vicinity of an impingement by a single
energetic monatomic ion or cluster ion. These studies show
significant deviations from the predictions of Sigmund's ellipsoidal
Gaussian form. For example, the molecular dynamics studies of Feix
{\it et al.} \cite{Feix05} indicate that for 5 keV ${\rm Cu^{+}}$
bombardment of Cu crystals, the collision cascade intensity along
the surface has a maximum along an annulus some distance from the
impact point and its spatial decay is better characterized by an
exponential rather than by a Gaussian function.  In this case Feix
{\it et al.} still found linear instability of a flat surface.
Moreover, in many cases \cite{Costantini01,Bringa01}, including low
energy (0.5 keV) bombardment of an amorphous silicon surface
\cite{Naga07}, the response of the surface is the formation of
craters with rims. This type of response, involving the accumulation
of matter at some locations, is in clear contradiction to the purely
erosive response predicted by Sigmund's model using a Gaussian
ellipsoid collision cascade. The occurrence of craters with rims has
been attributed to thermal spikes \cite{Bringa01} or to
ion-stimulated surface mass transport \cite{Naga07}.

These observations raise the interesting question of how robust are
the predictions of BH to the precise shape of the local response to
an ion impact. Indeed, the most general evolution equation based on
the accumulation of local responses to ion impacts is \cite{Aziz06}
\begin{equation}
\frac{\partial h({\bf x},t)}{\partial t} = \int {\bf dx'}
J_{ion}({\bf x'}) \Delta [{\bf x-x'}, h_{x}({\bf x},t), h_{y}({\bf
x},t), h_{xx}({\bf x},t), h_{yy}({\bf x},t), h_{xy}({\bf x},t),... ]
\ , \label{firsteq-aziz}
\end{equation}
where ${\bf x} = (x,y)$, $J_{ion}({\bf x'})$ is the ion flux at
${\bf x'}$, subscripts $x$ and $y$ denote partial derivatives, and
the kernel $\Delta [{\bf x-x'},\dots]$, representing the change in
height at ${\bf x}$ due to an ion impact at ${\bf x'}$, is expected
to decay smoothly to zero at large distances $|{\bf x} - {\bf x'}|$.
This equation is more general than that assumed by Sigmund because
the kernel $\Delta$ can have any shape whatsoever, and can depend on
the complete local geometry of the surface.

In this paper we explore whether a more general physically motivated
surface response can change the predictions for linear stability
from those of Bradley and Harper. Our purpose here is not to perform
quantitative comparison between theory and specific experiments, but
rather to determine how robust the predictions of the Bradley-Harper
theory are with respect to modifications of the ion impact function
$\Delta$. We demonstrate that, whereas the fundamental prediction
concerning the instability of flat surface to uniform ion
irradiation results from a wide class of response functions
including Gaussian and non-Gaussian distributions -- thus explaining
the applicability of Bradley-Harper theory for wide range of systems
-- there are certain classes of modification that have a dramatic
effect. Notably, these modifications render the flat surface stable
-- in contradiction to the classical theory -- while imperceptibly
affecting the yield curve $I(b)$.

The paper is organized as follows: In section {II} we extend the BH
approach -- of deriving from the microscopic response function the
coefficients $S_x(b),S_y(b)$ in Eq. (\ref{eq0a}) -- to a broad class
of purely erosive surface response functions, of which the Gaussian
ellipsoid is a particular example and the response of Feix {\it et
al.} \cite{Feix05} is another example. We show that the BH
prediction of linear surface instability for all incident beam
angles is unchanged. Hence any purely erosive surface response
within this broad class is contradicted by experiments. In the
remainder of the paper we explore possible physical mechanisms that
could  resolve this condundrum.  In section {III}, we demonstrate
that a surface response that is not purely erosive, but rather
consists of the formation of a crater surrounded by a rim, does
allow linear stability for some range of incidence angles. In
section {IV}, we demonstrate that impact-induced ``downhill''
surface currents, such as those recently found in MD simulation of C
and Si surfaces bombarded by low energy ($\sim 250$ keV) ions
\cite{Moseler05}, can also yield linear stability for some range of
beam angles. There are thus multiple physical mechanisms that could
explain the experiments, and the essential question is to determine
which effect is dominant. Identifying the dominant physical
mechanism for linear (in)stability is critical to having a reliable
nonlinear theory for pattern formation. In section V, we discuss how
experiments might distinguish the competing theories. In particular
we argue for a   careful analysis of experiments near the observed
critical angle at which a flat surface becomes stable.
\section{Bradley-Harper theory revisited}

The Sigmund theory of sputtering \cite{Sigmund69} posits that the
local erosion of the surface in Eq. (\ref{firsteq-aziz}), $\Delta
[{\bf x-x'}] / \sqrt{1+b^2}$ with $b$ the local surface slope, is
proportional to the local atom emission rate resulting from the
nuclear collision cascade, which itself is proportional to the
nuclear energy deposition density at $({\bf x}, h({\bf x}))$ from an
ion impinging at $({\bf x'}, h({\bf x'}))$. To demonstrate the source
of an instability\cite{Sigmund73}, Sigmund modeled the collision
cascade as a Gaussian ellipsoid. Bradley and Harper's subsequent
expansion of Sigmund's Gaussian ellipsoid collision cascade model,
combined with smoothening by fourth-order Mullins-Herring surface
diffusion, leads to Eq. (\ref{eq0a}).

To examine the consequences of forms of the erosive response that
are more general than Gaussian ellipsoids, we assume:
\begin{eqnarray}
\Delta [{\bf x-x'},\dots]  &=& \Delta h(r,z) \nonumber \\ &=& -{A}
e^{-g(r) - f(z)}  \ , \label{response}
\end{eqnarray}
where $r = \sqrt{x^2 + y^2}$, $z = h(x,y)$, and ${A}$ is a length
that depends on parameters such as ion energy and ion and target
mass. The first equality in Eq. (\ref{response}) assumes radial
symmetry about the ion track and no explicit dependence on the
surface slope and curvature, with the kernel depending only on $r$
and $z$. The second equality assumes separation of the variables $r$
and $z$. In Eq. (\ref{response}) the ion is assumed to penetrate the
surface at $(r,z)=(0,0)$.

Sigmund's Gaussian ellipsoid response is a particular case of Eq.
(\ref{response}), with
\begin{equation}
f(z) = \frac{1}{2\sigma^2} (z-a)^2 \ \ ; \ \ g(r) = \frac{1}{2 \mu^2}
r^2  \ , \label{sigmund}
\end{equation}
where $a$ is the average penetration depth of the ion, and $\sigma$, $\mu$
are lengths characterizing the ranges of response in directions
parallel and perpendicular to $\hat{z}$, respectively.

Following Bradley-Harper, we substitute in Eq. (\ref{firsteq-aziz})
the response form (\ref{response}) and add a relaxation mechanism to
the surface dynamics associated with Herring-Mullins surface
diffusion:
\begin{equation}
\frac{\partial h}{\partial t}  =- B \nabla^4 h  -\alpha
\int_{-\infty}^{\infty}\!\! dy \int_{-\infty}^{\infty} \!\! dx  \
e^{-g(\sqrt{x^2 + y^2}) - f(h(x,y))}  , \label{eq1}
\end{equation}
where $\alpha=A J_{ion}$ and the materials parameter $B$ is given by
$B=\gamma \Omega^{2}DC/(k_{B}T)$.  Here $C$, $D$, and $\Omega$ are
the concentration, diffusivity, and volume, respectively of the
surface-diffusing species; $\gamma$ is the surface free energy,
$k_B$ is Boltzmann's constant and $T$ is the absolute temperature.

To study evolution of surface morphology in the limit that the
surface height $h(x,y,t)$ varies on scales much larger than the ion
penetration depth, we consider perturbations about a planar surface
$(x,y,h=bx)$, so that
 $$h(x,y) = bx + \frac{1}{2} h_{xx} x^2 +
\frac{1}{2} h_{yy} y^2 + h_{xy} x y + \cdots \ ,  $$ and expand
$e^{-f(h(x,y))}$ to obtain
\begin{equation}
\exp [-f(h(x,y))] \approx e^{-f(bx)}[1 - f'(bx)[\frac{1}{2} h_{xx}
x^2 + \frac{1}{2} h_{yy} y^2 + h_{xy} x y]] \ . \label{eq2}
\end{equation}
With the expansion (\ref{eq2}), the integral equation (\ref{eq1}) is
readily transformed into the PDE (\ref{eq0a}) with the coefficients:
\begin{eqnarray}
I(b) &=& \alpha \int_{-\infty}^{\infty}\!\! dy \int_{-\infty}^{\infty}
\!\! dx  \ e^{-\rho_b(x,y)} \nonumber \\
S_y(b) &=& \alpha \int_{-\infty}^{\infty}\!\! dy \int_{-\infty}^{\infty}
\!\! dx  \ e^{-\rho_b(x,y)}  f'(bx)  y^2 \nonumber \label{sy} \\
S_x(b) &=& \alpha \int_{-\infty}^{\infty}\!\! dy \int_{-\infty}^{\infty}
\!\! dx  \ e^{-\rho_b(x,y)} f'(bx) x^2 \label{eq4}
\end{eqnarray}
where $\rho_b(x,y) = g(\sqrt{x^2 + y^2}) + f(bx)$.

The question now is how various choices of $f(r)$ and $g(z)$ can
change $I(b) , \  S_x(b)$ and $S_y(b)$. We are primarily interested
in the slope dependence in $S_y(b)$, because in the Bradley-Harper
theory $S_y(b)<0$ for all slopes $b$. Our question is whether any
choice of $f(z),g(r)$ can stabilize the surface against
perpendicular modes ($S_y>0$) for some range of $b$ while not
significantly affecting the shape of the yield curve.  The latter
requirement is especially significant because the yield curve
predicted by the Sigmund response function agrees qualitatively with
that measured on many materials -- at least for non-grazing
incidence \cite{vasile99}.

All of our analysis proceeds with the same methodology: the integral
for $S_y(b)$ in Eq. (\ref{sy}) is dominated by contributions near the
minimum of $\rho_b$ which we call $\{x_{min},y_{min}\}$.  This is
because the size of the region where energy is deposited (of order
the penetration depth $a$) is much smaller than the characteristic
length scale over which the surface shape varies. The minima of
$\rho_b$ satisfy the  equations
\begin{eqnarray}
\frac{y_{min}}{\sqrt{x_{min}^2 + y_{min}^2}} g'(\sqrt{x_{min}^2 + y_{min}^2}) &=& 0; \\
\frac{x_{min}}{\sqrt{x_{min}^2 + y_{min}^2}} g'(\sqrt{x_{min}^2 +
y_{min}^2}) + b f'(b x_{min}) &=& 0.
\end{eqnarray}
Depending on the functional forms of $g$ and $f$ there are two
possible types of solutions to these equations.
\begin{eqnarray}
&{\emph (a)}&  \ \ y_{min} = 0 , \ \pm g'(x_{min}) + b f'(b x_{min})
=
0 \label{fp_a}\\
&{\emph (b)}& g'(\sqrt{x_{min}^2 + y_{min}^2}) = 0 , \ b f'(b
x_{min}) = 0  \ , \label{fp_b}
\end{eqnarray}
where the $\pm$ signs in (\ref{fp_a}) correspond to $x_{min} > 0,
x_{min}<0$, respectively. Once the locations of the minima are
determined, we can expand
\begin{eqnarray}
\rho_b&=&\rho_b(x_{min},y_{min}) + \frac{(x-x_{min})^2}{2} (g_{xx} +
b^2 f'') \nonumber \\ &+& (x-x_{min})(y-y_{min}) g_{xy} +
\frac{(y-y_{min})^2}{2} g_{yy} \nonumber \\ &\equiv& \rho^* + \tilde
A (x-x_{min})^2 + \tilde C(x-x_{min})(y-y_{min}) + \tilde B
(y-y_{min})^2 \label{expa}
\end{eqnarray}
where the second equality defines $\tilde A,\tilde B,\tilde
C,\rho^*$. This expansion can then be used to evaluate the integral.

We now proceed to use this methodology to establish the conclusion
that $S_y\le 0$ is extremely robust. For any kernel of the form
considered here a perpendicular mode instability always exists for
all slopes $b$. The characteristic behavior of the coefficient $S_x$
is more fickle. Obviously, for $b\to 0$, $S_x(b)/S_y(b) \to 1$, and
therefore $S_x(b)$ is necessarily negative for small enough slopes
$b$. However, Bradley-Harper's observation, that Gaussian ellipsoids
imply $S_x <S_y <0$ for $b\ll 1$, does depend on the exact shape of
the response function. This can be readily verified by considering
Sigmund's response Eq. (\ref{sigmund}) with $a<\sigma$. Hence, we
will focus our analysis on the robust properties of the linear
dynamics, associated with the sign of $S_y$, and will not further
discuss $S_x$ in this section.

\subsection{The shape of the energy distribution does not qualitatively affect stability}

We begin by considering changes in only the {\sl shape} of the
energy distribution: namely we consider $f(z),g(r)$ that keep the
position of maximum energy deposition at a single point (the average
stopping point of the ion), though we vary the shape of the
distribution. We thus assume that the function $f(z)$ has a minimum
at $z=a$ whereas $g(r)$ increases monotonically from $r=0$.

Under these assumptions, the minimum of $\rho_b(x,y)$ must be of
type (\ref{fp_a}). Moreover, because the minimum of $g(r)$ along the
$x$ axis occurs at $x=0$ and the minimum of $f(bx)$ occurs at
$x=a/b>0$, then $x_{min}$, determined by $g'(x_{min})+ b f'(b
x_{min})=0$ must be in the interval $0<x_{min} < a/b$, such that
$f'(bx_{min}) < 0$. The expansion of $\rho_b$ in equation
(\ref{expa}) leads to the coefficients $\tilde A=(g''+b^2 f'')/2$,
$\tilde B=g'/2 |x_{min}|$ and $\tilde C=0$, where all derivatives
are taken at $x_{min}$. Hence the integral is approximately
\begin{equation}
S_y(b) \approx \alpha \int_{-\infty}^{\infty}\!\! dy
\int_{-\infty}^{\infty} \!\! dx  \ e^{-\rho^*-\tilde A
(x-x_{min})^2-\tilde B y^2} f'(bx_{min})  y^2.\label{syappr}
\end{equation}
Because $f'(bx_{min})<0$ the integral (\ref{syappr}) is necessarily
negative for all $b$. This demonstrates that the experimentally
observed stability of a sputtered surface to perpendicular mode
ripples is not a consequence of the shape of the energy
distribution.

\subsection{Toroidal energy distributions do not qualitatively affect stability}
Another possible modification of the energy distribution is for the
maximum energy deposition to occur away from the ion trajectory.
Indeed, Feix \textit{et al.}'s recent simulations of Cu crystals
bombarded by 5 keV ${\rm Cu^+}$ ions \cite{Feix05} have demonstrated
energy distributions with a maximum along an annulus surrounding the
ion trajectory. Such a response is thus characterized by a $g(r)$
with a minimum at $r_{min}=r_0 >0$.

Consider the sign of $S_y(b)$ under these circumstances. There are
now two different regimes, depending on the slope.  When the slope is
small, such that $a/b\ge r_0$, the minimum must be of type (a), Eq.
(\ref{fp_a}). Type (b) (Eq. (\ref{fp_b})) is excluded because if
$f'(bx_{min})=0$ then we must have $x_{min}=a/b$. But then the
equation $g'(r)=0$ cannot be satisfied: this equation implies that
$x_{min}^2+y_{min}^2=r_0^2$, which cannot be obeyed for any
$y_{min}$. In contrast, when the slope is large, so that $a/b\le
r_0$, the minima are of type (b).

Let us first consider the regime of small slope. Here the analysis
proceeds as above with the same $\tilde A,\tilde B,\tilde C$ defined
in (\ref{expa}). As before the sign of the integral hinges on the
value of $f'(bx_{min})=-g'(x_{min})/b$. Because we are assuming that
the minimum of $f(bx)$ occurs at $x=a/b$ which is larger than the
minimum assumed by $g(r)$ along the $x$-axis, at $x=r_0$, Eq.
(\ref{fp_a}) implies that $f'(bx_{min}) < 0$. Hence we arrive at the
conclusion that in the small slope regime $S_y \le 0$: the linear
instability survives.

The second regime,  where $b/a\le r_0$,  is more subtle, with two
minima being of type (b) (Eq. \ref{fp_b}). Assuming the minimum of
$g(r)$ occurs at $r_0$, and the minimum of $f(z)$ occurs at $a$, in
this case we have that
$(x_{min},y_{min}^{\pm})=(a/b,\pm\sqrt{r_0^2-(a/b)^2}).$ The value
of $S_y(b)$ is given by the sum of the contributions to the integral
centered around  each of these two minima. For these minima the
values of $\tilde A,\tilde B,\tilde C$ are given by
 \begin{equation}
 \tilde A=\frac{1}{2}\biggl ( g'' \frac{x_{min}^2}{r_0^2} + b^2 f''\biggr) \
 , \ \tilde B=\frac{1}{2}  g'' \frac{y_{min}^2}{r_0^2}  \ , \
 \tilde C^{\pm}= g'' \frac{x_{min}y_{min}^{\pm}}{r_0^2} \ ,
 \label{notations1}
 \end{equation}
 where $g''$ is evaluated at $r=r_0$ and $f''$ is evaluated at $z=a$.
 We now must evaluate
 \begin{equation}
S_y(b) \approx \alpha \sum_{\pm}\int_{-\infty}^{\infty}\!\! dy
\int_{-\infty}^{\infty} \!\! dx  \ e^{-\rho^*-\tilde A
(x-x_{min})^2-\tilde B (y-y_{min}^{\pm})^2-\tilde C^{\pm}
(x-x_{min})(y-y_{min}^{\pm})} f'(bx) y^2.
\end{equation}
The exponential in the integrals are best dealt with by completing
the square, so that they become
 \begin{equation}
S_y(b) \approx \alpha \sum_{\pm} \int_{-\infty}^{\infty}\!\! dy
\int_{-\infty}^{\infty} \!\! dx  \ e^{-\rho^*-\tilde A\biggl(
(x-x_{min})+ \frac{\tilde C^{\pm}}{2\tilde A}
(y-y_{min}^{\pm})\biggr)^2 } e^{- (y-y_{min}^{\pm})^2(\tilde
B-(\tilde C^{\pm})^2/4\tilde A)} f'(bx)  y^2.
\end{equation}
Now the second exponential decays with $y$ varying away from
$y_{min}^{\pm}$ because $\tilde B\ge \frac{(\tilde
C^{\pm})^2}{4\tilde A}$ for any $b \neq 0$. If we now change
variables to $\tix=x-x_{min} + \frac{\tilde C^{\pm}}{2\tilde A}
(y-y_{min}^{\pm})$ and $\tiy =y-y_{min}^{\pm}$ we obtain
 \begin{equation}
S_y(b) \approx e^{-\rho^*} \alpha \sum_{\pm}
\int_{-\infty}^{\infty}\!\! d\tiy \int_{-\infty}^{\infty} \!\! d\tix
\ e^{-\tilde A \tix^2 - \tiy^2 (\tilde B-(\tilde C^{\pm})^2/4\tilde
A)} f'[b( x_{min}+\tix - \frac{\tilde C^{\pm}}{2\tilde A} \tiy)]
(y_{min}^{\pm}+\tiy)^2 \ .
\end{equation}
Because now $f'(bx_{min})=0$, evaluation of the integrals to leading
order requires expansion of the terms $f'[b( x_{min}+\tix
-\frac{\tilde C^{\pm}}{2\tilde A} \tiy)]$ around $a=bx_{min}$. With
this we get the following approximation to the integral:
\begin{equation}
S_y(b) \approx e^{-\rho^*} \alpha \sum_{\pm}
\int_{-\infty}^{\infty}\!\! d\tiy \int_{-\infty}^{\infty} \!\! d\tix
\ e^{-\tilde A \tix^2 - \tiy^2 (\tilde B-(\tilde C^{\pm})^2/4\tilde
A)} \biggl(f''(a) (b \tix-\frac{\tilde C^{\pm}}{2\tilde A} b \tiy)+
\cdots \biggr) (y_{min}+\tiy)^2.
\end{equation}
The contribution of the two integrals is identical, and sums up to:
\begin{equation}
S_y(b)\approx -f''(a)g''(r_0) \frac{a y_{min}^2}{r_0^2 \tilde A}
\Gamma_1 \label{result}
\end{equation}
where we have substituted the formula for $C^{\pm}$
(\ref{notations1}), have used $bx_{min}=a$, and where
$$\Gamma_1= e^{-\rho^*} \alpha \int_{-\infty}^{\infty}\!\! d\tiy
\int_{-\infty}^{\infty} d\tix e^{-\tilde A \tix^2}e^{- \tiy^2
(\tilde B-\tilde C^2/4\tilde A)} \tiy^2  \ . $$ The RHS of Eq.
(\ref{result}) is negative definite, i.e. $S_y(b)$ is negative for
all values of $b$. Hence response functions of the form of Eq.
(\ref{response}) generally cause a perpendicular mode instability
for any incidence angle. The qualitative conclusions of the original
Bradley-Harper analysis concerning the instability of perpendicular
surface modulations at any beam angle are thus very robust.


\section{Effects of mass redistribution}

The analysis of the previous section demonstrates that a broad class
of purely erosive response functions gives rise to linear instability for
all beam angles.  However, there have been several recent studies
suggesting that the surface response is not purely erosive. These
studies demonstrate that after ion impact, a crater forms around the
impact point of the penetrating ion, surrounded by rims elevated
from the original surface \cite{Bringa01,Yamada01,Naga07,Costantini01} . This behavior, where $\Delta h > 0$ in
the rim, is completely different from the erosive response functions
described above. We investigate whether such response functions can
cause the stability of a flat surface.

To carry out this analysis,  we introduce a natural
generalization of the family of response functions (\ref{response}):
\begin{equation}
\Delta h(r,z) = -\sum A_j e^{-g_j(r) - f_j(z)} \label{responsemod}
\end{equation}
where $g_j(r),f_j(z)$ are localized functions as discussed in the
previous section, but the coefficients $A_j$ can be negative or
positive. In particular, negative $A_j$ corresponds to mass
deposition associated with ion impact and can give rise to formation
of rims. A particularly simple form of a response function is the sum
of two Gaussian ellipsoids:
\begin{equation}
\Delta h(r,z) =-A [ e^{-r^2/2\mu_1^2 - (z-a_1)^2/2\sigma_1^2} -
\beta e^{-r^2/2\mu_2^2 - (z-a_2)^2/2\sigma_2^2} ] \ .
\label{Gaussians2}
\end{equation}
This response function has eight free parameters (including $A$ and
$\beta$), all of which are constrained to be positive.  Unlike the
original Sigmund model, the free parameters here are not directly
connected to a microscopic picture. Because our intent is to
understand whether small deviations from Sigmund's response function
can change the stability characteristics of the surface, we will
consider the case with $\beta\ll 1$, and think of
$a_1,\mu_1,\sigma_1$ as corresponding essentially to the original
Sigmund parameters. The parameters $a_2,\mu_2,\sigma_2$ describe
characteristics of the mass redistribution.

With the model so defined, we can evaluate the yield curve $I(b)$ as
well as $S_x(b), S_y(b)$, obtaining
\begin{eqnarray}
I(b) &=& 2 \pi J_{ion}\sum_{i=1,2} A_i \mu_i^2 \sigma_i \sqrt{\frac{1 +
b^2}{b^2 \mu_i^2 + \sigma_i^2}} e^{-a_i^2 / [2 (b^2 \mu_i^2 +
\sigma_i^2)]} \label{2g-i} \\
S_x(b) &=& - 2 \pi J_{ion}\sum_{i=1,2} A_i a_i \mu_i^4 \sigma_i [b^2 a_i^2
\mu_i^2 - 2  b^4 \mu_i^4 - b^2 \mu_i^2 \sigma_i^2 + \sigma_i^4]
\sqrt{\frac{1 + b^2}{(b^2 \mu_i^2 + \sigma_i^2)^7}} e^{-a_i^2 / [2
(b^2 \mu_i^2 + \sigma_i^2)]}
\label{2g-sx}\\
S_y(b) &=& - 2 \pi J_{ion}\sum_{i=1,2} A_i a_i \mu_i^4 \sigma_i
\sqrt{\frac{1 + b^2}{(b^2 \mu_i^2 + \sigma_i^2)^3}} e^{-a_i^2 / [2
(b^2 \mu_i^2 + \sigma_i^2)]} \ ,
 \label{2g-sy}
\end{eqnarray}
where we used the notation $A_1=A, A_2=-\beta A$.

We now want to use this result to address the following question: is
there a regime of parameter space where the stability
characteristics of the surface are qualitatively different from the
predictions of Bradley and Harper, but for which the yield curve is
experimentally indistinguishable from that predicted by the Sigmund
response? Indeed,
we have found multiple regions of parameter space where this occurs.  This can be
demonstrated simply and analytically by expanding equations
(\ref{2g-i},\ref{2g-sx},\ref{2g-sy}) in the regime of small slopes,
where $S_x \approx S_y$. We find that, as $b\to 0$,
\begin{eqnarray}
I(b) &\approx& 2\pi J_{ion}A[\mu_1^2 e^{-a_1^2/(2 \sigma_1^2)}-
\beta \mu_2^2
e^{-a_2^2/(2 \sigma_2^2)}] \\
S_y(b) &\approx& S_x(b) \approx -2\pi J_{ion} A
[\frac{a_1\mu_1^4}{\sigma_1^2}e^{-a_1^2/(2 \sigma_1^2)} - \beta
\frac{a_2\mu_2^4}{\sigma_2^2}e^{-a_2^2/(2 \sigma_2^2)}] \  ,
\end{eqnarray}
Here we see that for small slopes, $S_x$ and $S_y$ can have either
sign, depending on the relative magnitudes of the terms
$\frac{a_1\mu_1^4}{\sigma_1^2}e^{-a_1^2/(2 \sigma_1^2)} $ and
$\beta\frac{a_2\mu_2^4}{\sigma_2^2}e^{-a_2^2/(2 \sigma_2^2)}$. If
the second term dominates the first then $S_x$ and $S_y$ are
positive at small $b$ and the surface is stable to all
perturbations. Can stability be achieved without significantly
affecting $I(b)$? Obviously, this will be the case if $\mu_1^2
e^{-a_1^2/(2 \sigma_1^2)} \gg \beta \mu_2^2 e^{-a_2^2/(2
\sigma_2^2)}$. Letting $Z_i=\mu_i^2 e^{-a_i^2/(2\sigma_i^2)}$,
satisfaction of the two conditions amounts to finding parameters
where (i) $ Z_a a_1\mu_1^2/\sigma_1^2  < \beta Z_2 a_2
\mu_2^2/\sigma_2^2 $ while (ii) $Z_1/Z_2\gg\beta$.  We also would
like $\beta$ to remain small. Such a parameter regime clearly exists
and merely constrains the scale and the geometry of the mass
redistribution region.

\begin{figure}
\begin{center}
\includegraphics[width=2.0in,clip=]{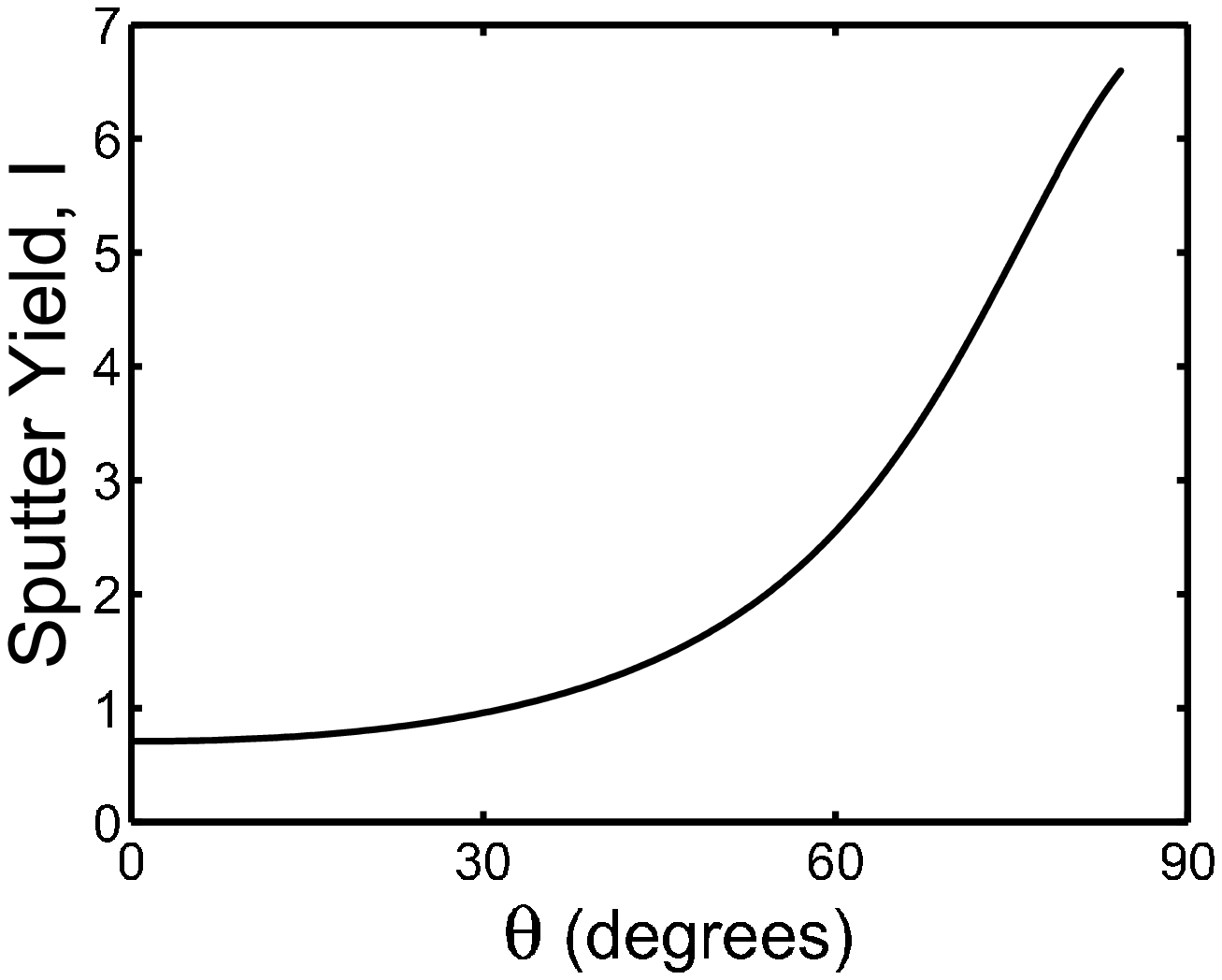}
\includegraphics[width=2.0in,clip=]{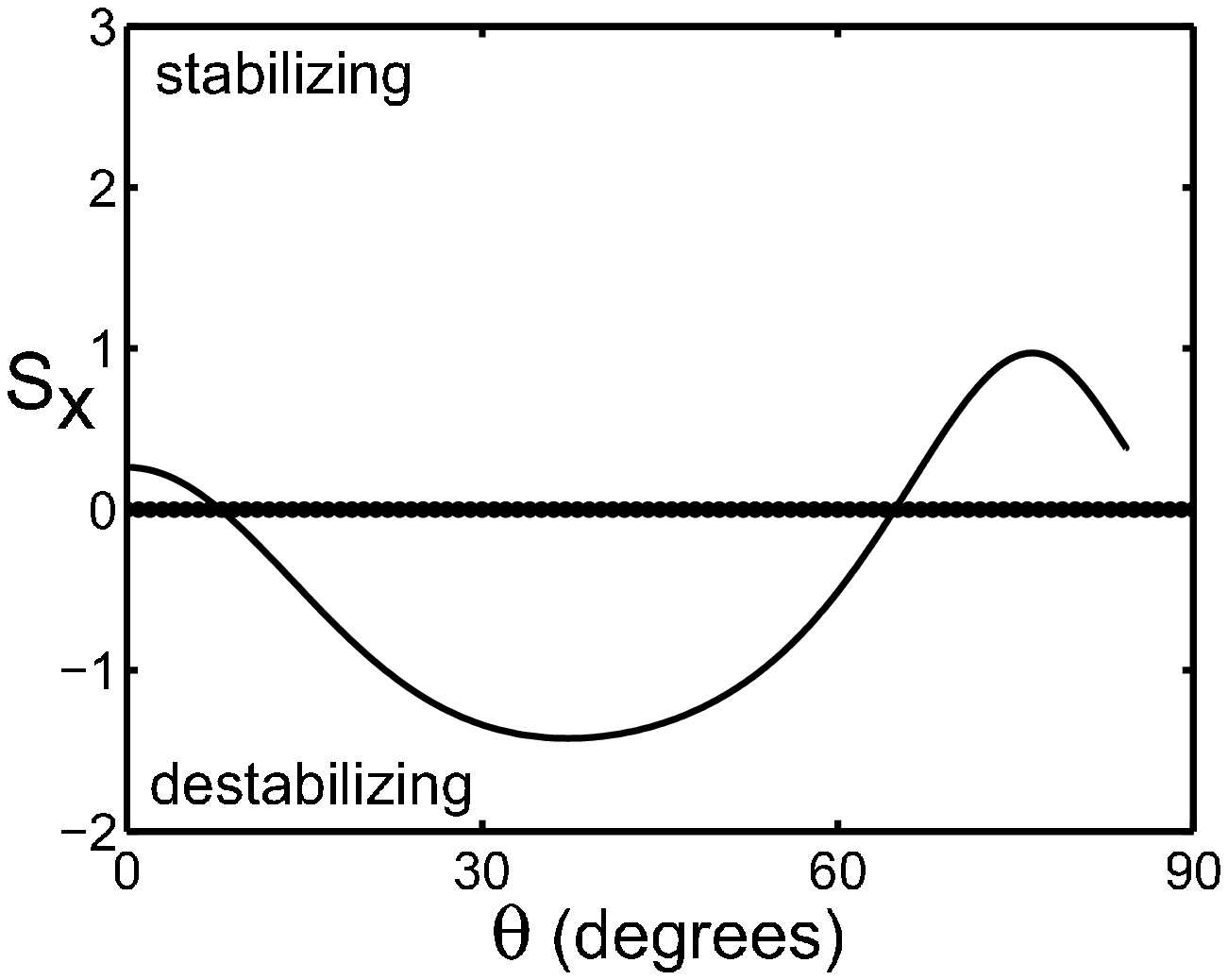}
\includegraphics[width=2.0in,clip=]{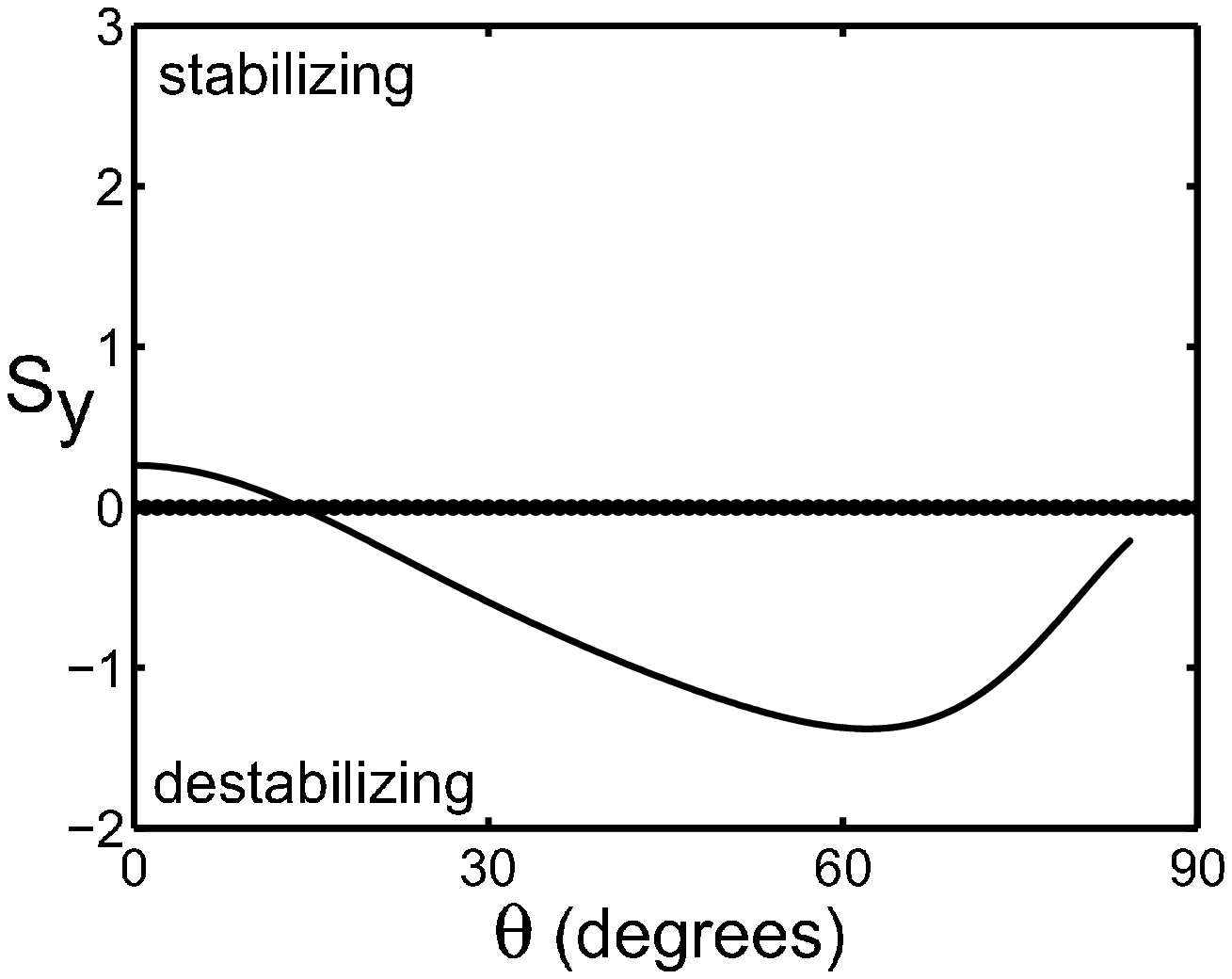}
\includegraphics[width=2.0in,clip=]{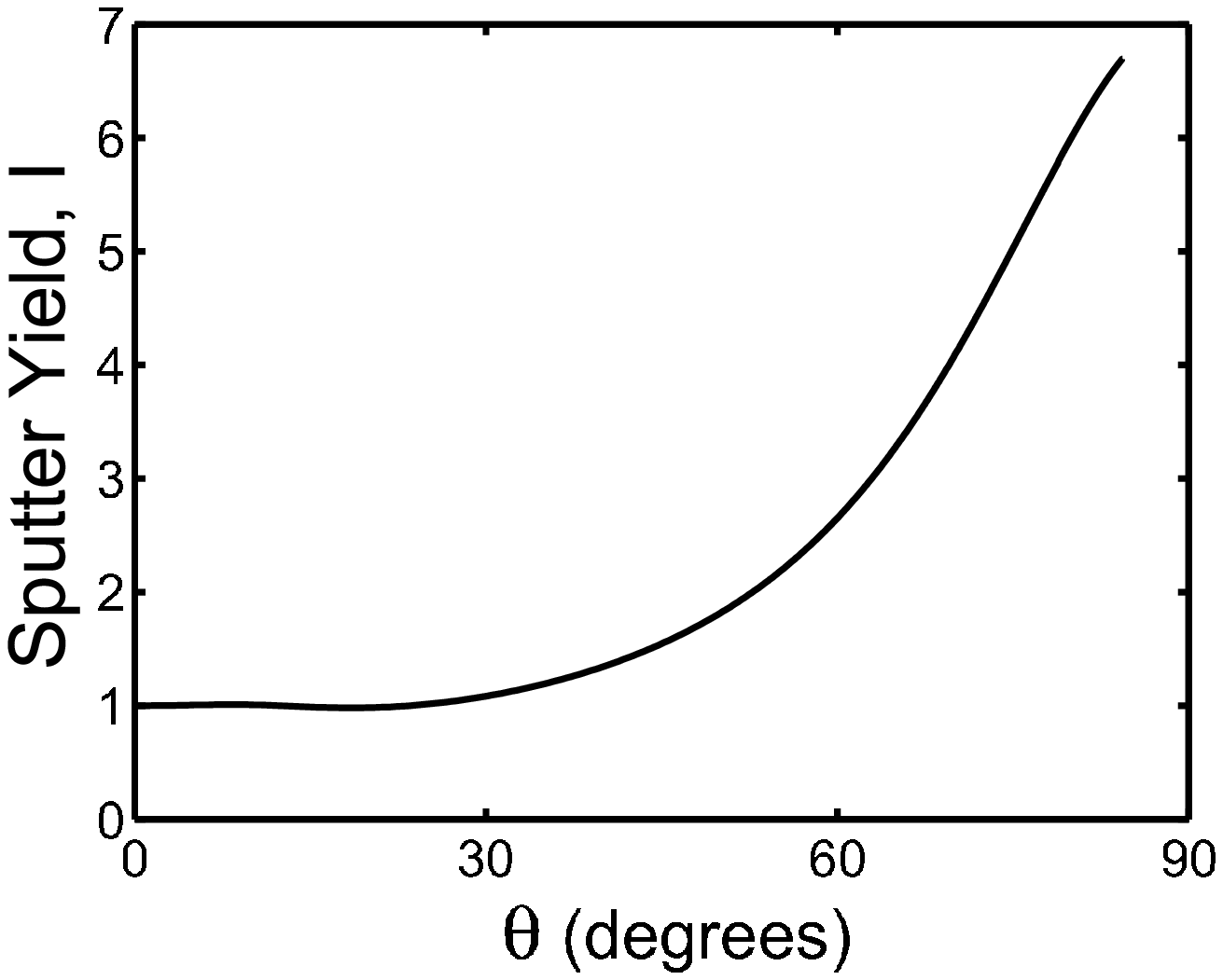}
\includegraphics[width=2.0in,clip=]{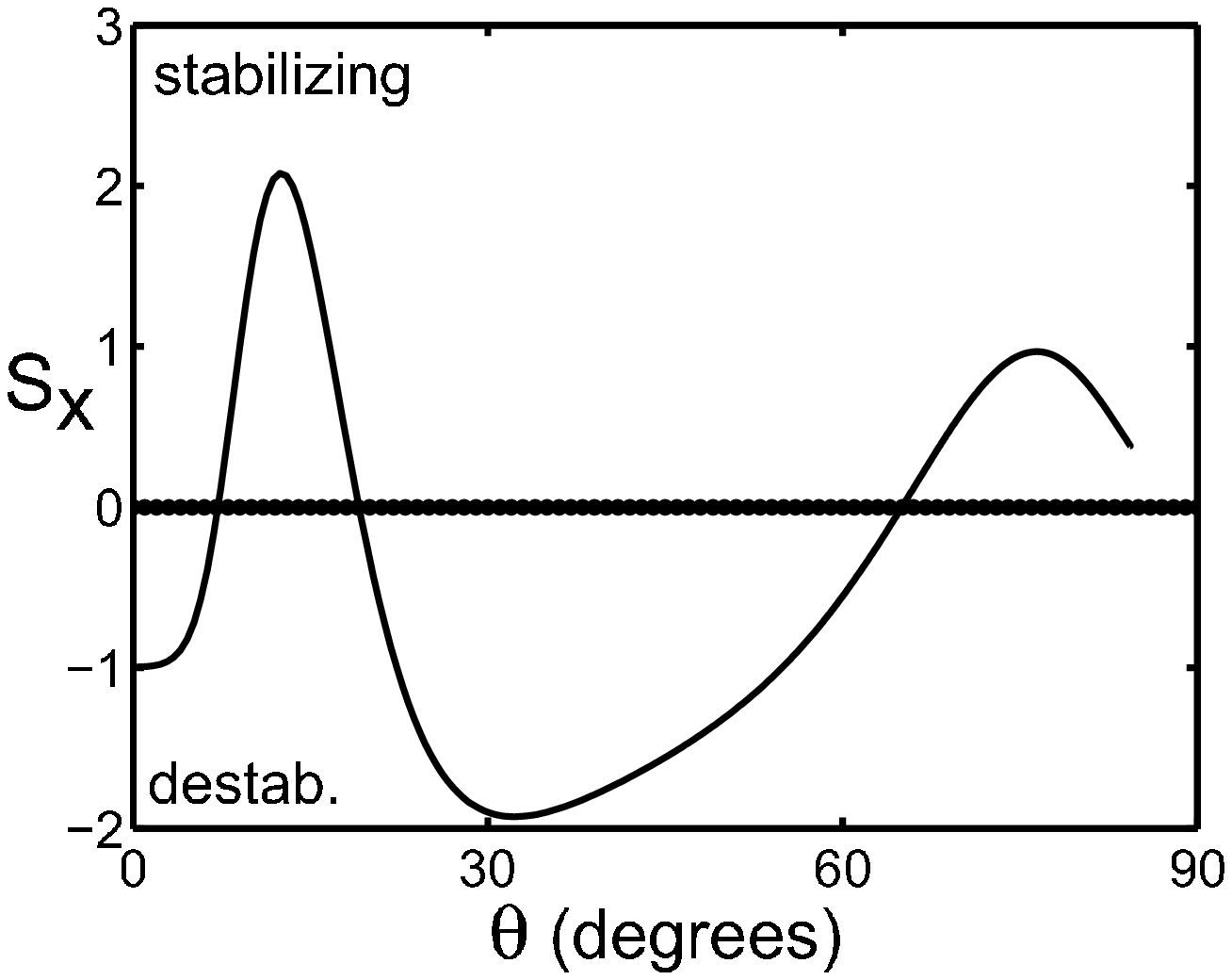}
\includegraphics[width=2.0in,clip=]{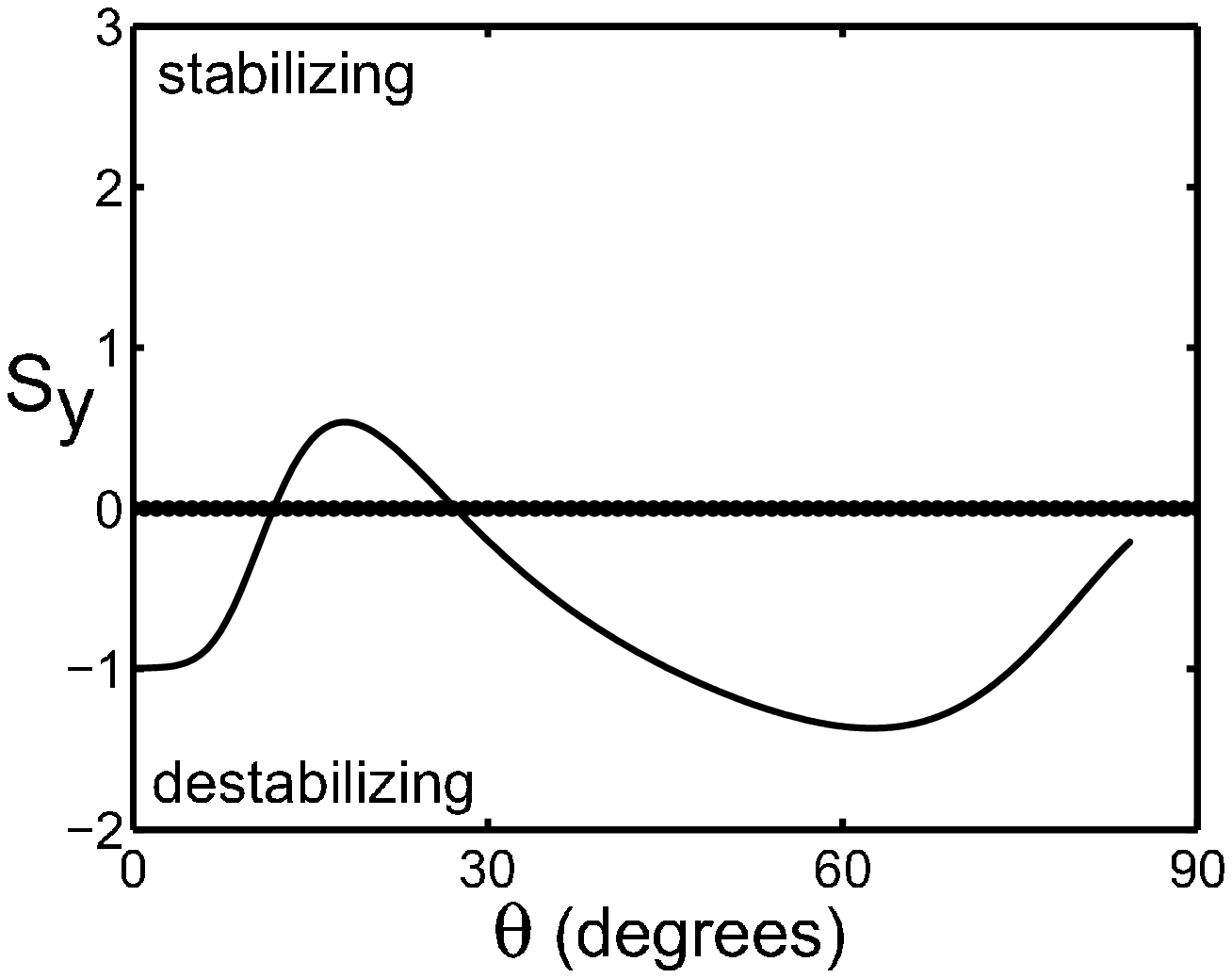}
\caption{\label{figTwoGauss} Normalized yield curve and BH
coefficients $S_x,S_y$ for two sets of parameters of the
two-Gaussians model, Eq. (\ref{Gaussians2}). The "Sigmund
parameters" $a,\sigma,\mu$ are taken as in Fig. \ref{figBH}, and the
same normalization factors are used. The new parameters are: (a)
$\alpha = 0.03; a_2 = 0.5 \textrm{nm,} \sigma_2 = 0.5$ nm, and
$\mu_2 = 1$ nm and (b) $\alpha=0.03$, $a_2=0.9\textrm{nm,}$
$\sigma_2=0.2$ nm and $\mu_2=1.5$ nm.}
\end{center}
\end{figure}

 To demonstrate this explicitly, Fig.
\ref{figTwoGauss} shows the behavior of $I(b),S_x(b), S_y(b)$, where
we have used the same parameters for
$a_1=1.5,\sigma_1=0.9,\mu_1=0.5$ as used for the 'normal'
Bradley-Harper stability characteristics shown in Fig. \ref{figBH},
with the additional parameters $\beta = 0.03,  \ a_2 = 0.5
\textrm{nm,} \ \sigma_2 = 0.5$ nm, and $\mu_2 = 1$ nm. For these
parameters $Z_1=0.06 \textrm{nm}^2$ and $Z_2 = 0.6 \textrm{nm}^2$,
thus $Z_1/Z_2 \gg \beta$, whereas $a_1\mu_1^2/\sigma_1^2=0.46$ nm
and $a_2 \mu_2^2/\sigma_2^2=2$ nm. We therefore satisfy both
constraints (i) and (ii) listed above. Indeed, the top row of Fig. 2
shows a stable region of parameter space for small slopes in {\sl
both} $S_x$ and $S_y$, while the qualitative shape of the yield
curve is unchanged.
We have also found regions of parameter space where the two
conditions derived above are not met, hence a flat surface is
unstable at small $b$, but there is still a window of stability at
higher slopes, as shown in the bottom row of Fig. 2.

The results of this section demonstrate a very significant
conclusion: that small changes in the shape of the surface response
of a single ion can completely change the stability characteristics
of a flat surface from those predicted by Bradley and Harper, but
yet not lead to any significant modification to the measured yield
curve. Further analysis along this line requires a microscopic
theory for the non-erosive processes, or detailed atomistic
simulations from which effective parameters such as $\beta,
a',\sigma'$, and $\mu'$ can be determined.


\section{Induced surface currents}
In the previous sections we considered a surface response that does
not depend explicitly on the incidence angle and is fully
characterized by considering normal incidence ($b=0$). Namely, the
response at a point $(x,y,h(x,y))$ depends only on the projections
of the vector that connects $(x,y,h(x,y))$ to the average ion
stopping point $(0,0,-a)$, in directions parallel and perpendicular
to the beam direction $\hat{z}$.  Thus, the dependence of the
coefficients $I(b),S_x(b)$ and $S_y(b)$ in Eq. (\ref{eq0a})  on the
angle $\theta = \tan^{-1} (b)$ is implicit and purely geometrical,
stemming from the fact that the distribution of values of these
projections ($|bx+a| \ , \sqrt{x^2+y^2}$, respectively) over all
surface points depends on the slope $b$.

It is possible, however, that the response of a surface point to ion
impact depends explicitly on the incidence angle. Such behavior was
reported by Moseler {\it et al.} \cite{Moseler05}, who used molecular
dynamics to study the ion-enhanced smoothening of diamond-like carbon
surfaces bombarded by low energy (30-150 eV) carbon ions. These
authors simulated surfaces tilted at angles up to $20^o$ and observed
transient surface currents with components along the projection of
the ion beam direction onto the surface, resulting in net
displacements along the surface of magnitude proportional to the
incidence angle.  Their analysis of this effect, neglecting
densification and sputter erosion, and focusing on beam angles near
normal incidence, resulted in an isotropic diffusion-like equation
for the surface height:
\begin{equation}
\partial h /\partial t = \nu \nabla^2 h \ ,
\label{Moseler-term}
\end{equation}
where $\nu$ is positive and consequently stabilizing (c.f. Eq. (1)).
Moseler {\it et al.} did not pursue the beam angle dependence of
$\nu$. As is the case for the erosion coefficients in Eq.
(\ref{eq0a}), we expect this smoothening effect to become anisotropic
away from normal incidence, yielding two different coefficients
$\nu_x(b)$, $\nu_y(b)$.

Previously, Carter and Vishnyakov \cite{Carter96} proposed a similar
smoothening term to explain the absence of linear instability on
silicon bombarded with 10-40 keV ${\rm Xe^{+}}$ \textrm{at incidence
angles between 0 and } $45^{o}$. They proposed a mechanism whereby
forward recoils move, on average, parallel to the ion beam before
coming to rest.  They retained the projection along the surface,
which may be interpreted as a consequence of the incompressibility
of the solid: the surplus density injected into the solid
subsequently ``pops up'' to the surface along, on average, the
shortest path. Specifically, for an ion flux of magnitude $J_{ion}$
in a plane perpendicular to the ion beam, the number of ion
impingements per unit area of surface is $J_{ion} \cos(\theta)$,
where $\theta$ is the local angle of incidence. The induced current
per ion projected along the surface varies as $\sin(\theta)$,
resulting in a surface current $J_{x}$ proportional to $J_{ion}
\sin(\theta) \cos(\theta)$, or $J_{ion} \sin(2\theta)$. This surface
current has the same stabilizing effect on parallel mode
instabilities as that identified in the simulations of Moseler
\textit{et al.}, Eq. (\ref{Moseler-term}), but with $\nu_x \propto
\cos(2\theta)$. Carter and Vishnyakov did not consider $\nu_y$.

In principle, the low-energy mechanism of Moseler {\it et al.}
differs from the high-energy Carter-Vishnyakov mechanism: in the
former case, the projected range is $\sim 1$ nm and true surface
transport is observed; in the latter case, the projected range is
greater than 10 nm, volume transport is induced, and it is the
component parallel to the surface that results in the smoothening
effect. However, in both cases an explicit dependence on angle of
incidence is apparent, and phenomenologically they appear virtually
indistinguishable.  In both mechanisms the average net effect of
each ion impact is a displacement along the surface that is
proportional to $\theta$ for small $\theta$ and should saturate at
large $\theta$, as does $\sin(\theta)$. In all cases the ion
impingement rate per unit area of actual surface goes as
$\cos(\theta)$. Their combination should result in an induced
``downhill'' surface current that approaches zero near normal and
grazing incidence and displays a maximum in the vicinity of $45^o$.

To understand the implications of Eq. (\ref{Moseler-term}) for
linear stability, it is essential to establish the dependence on
incidence angle of both coefficients $\nu_x(\theta),\nu_y(\theta)$
for parallel and perpendicular modes, respectively. To this end we
consider a simple model in the spirit of those discussed above. The
geometry of the previous sections is assumed, where an ion flux
$J_{ion}$ impinges in the $-\hat{z}$ direction on a surface slightly
perturbed from the plane $h(x,y) = bx$, and $\theta$ is the angle
between the local normal to the surface and the $\hat{z}$ axis. We
assume that the component of ion momentum parallel to the surface
causes the displacement of surface target atoms a distance along the
surface proportional to $\sin(\theta)$. The contribution of the
induced surface current ${\bf J_s} = (J_x,J_y)$ to $\partial
h(x,y,t)/\partial t$ is $-\nabla \cdot {\bf J_{s}}$, where $\nabla =
(\partial_x ,\partial_y)$. In order to evaluate ${\bf J_s}$ let us
assume first that the surface is exactly described by  $h(x,y) =
bx$, where $b=tan(\theta)$. In this case $J_y = 0$, and with a
momentum component parallel to the surface proportional to
$sin(\theta)$, we obtain $J_x \propto - J_{ion} \cos(\theta)
\sin(\theta)$, where $J_{ion} \cos(\theta)$ is the rate of ion
impingement per unit surface area. This behavior is consistent with
the results of the MD simulations of Moseler et al.
(\cite{Moseler05}). In order to write the induced surface flux for a
general surface, represented by the equation $z = h(x,y)$, we must
express $J_x$ and $J_y$ in terms of $\nabla h$. The angle $\theta$
satisfies the relation $\cos(\theta) = \hat{n} \cdot \hat{z} =
1/\sqrt{|\nabla h|^2+1}$, where $\hat{n} = [-\partial h/\partial x,
-\partial h/\partial y,1]/\sqrt{|\nabla h|^2+1}$ is the unit vector
normal to the surface. Let us denote by $\phi$ the angle between $x$
axis and the direction within the $x-y$ plane of maximal increase in
surface elevation at $(x,y)$: $\phi = \tan^{-1} \frac{\partial
h/\partial y}{\partial h /
\partial x}$. The fluxes $J_x,J_y$ are then given by $J_x \propto
-\sin(2\theta) \cos(\phi)$ and $J_y \propto -\sin(2\theta)
\sin(\phi)$.

Because our analysis in this paper is restricted to linear dynamics
of the surface, we expand $\nabla \cdot J$ to linear order in
deviations of $h$ from the flat surface $h = bx$ ($b \neq 0$).
Algebraic manipulation yields the relations:
\begin{eqnarray}
\cos (\phi) \approx 1 \ &,& \  \sin (\phi) \approx  b^{-1}
\frac{\partial h}{\partial y} \ , \nonumber \\
\cos(\theta) \approx (1+b^2)^{-1/2} (1 - \frac{b}{1+b^2}
\frac{\partial h}{\partial x}) \ &,& \ \sin(2\theta) \approx
\frac{b}{b^2+1}[1+\frac{1-b^2}{b(1+b^2)}
\partial h /\partial x]\ ,
\end{eqnarray}
and the linear contributions $\nu_x(b),  \nu_y(b) $from the surface
induced currents to the coefficients $S_x(b), S_y(b)$, respectively,
in Eq. (\ref{eq0a}) is:

\begin{eqnarray}
&\nu_x(b)& \propto \frac{1-b^2}{(1+b^2)^2} \\
 &\nu_y(b) & \propto \frac{1}{1+b^2} \label{contribution}
\end{eqnarray}

The expression for $\nu_x $ in Eq. (\ref{contribution}) is
equivalent to the expression derived by Carter and Vishnyakov
\footnote{Note, however, that Carter and Vishnyakov used a different
coordinate system, in which the $\hat{x}$ direction is parallel to
the surface.}.
\begin{figure}
\begin{center}
\includegraphics[width=2.6in,clip=]{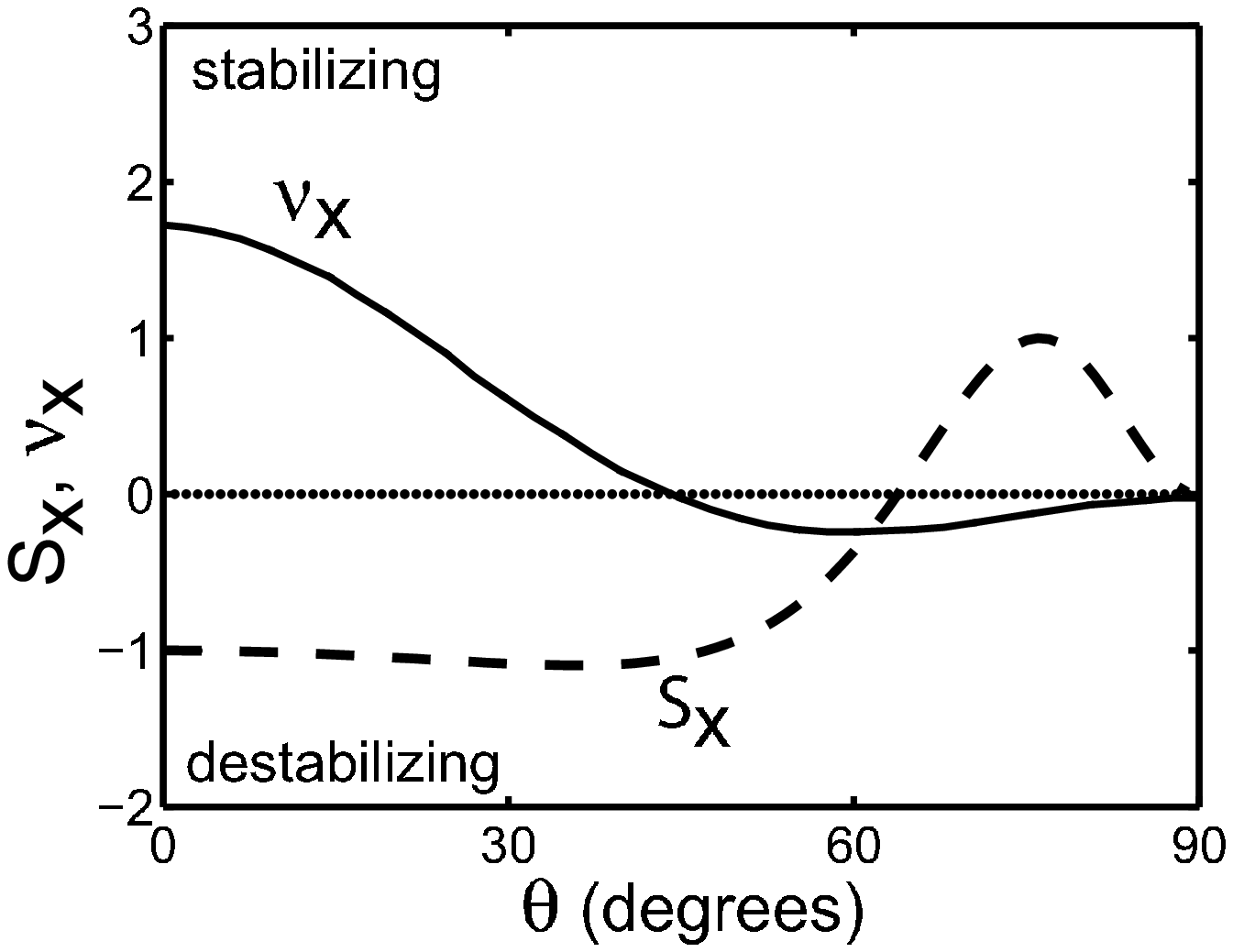}
\includegraphics[width=2.6in,clip=]{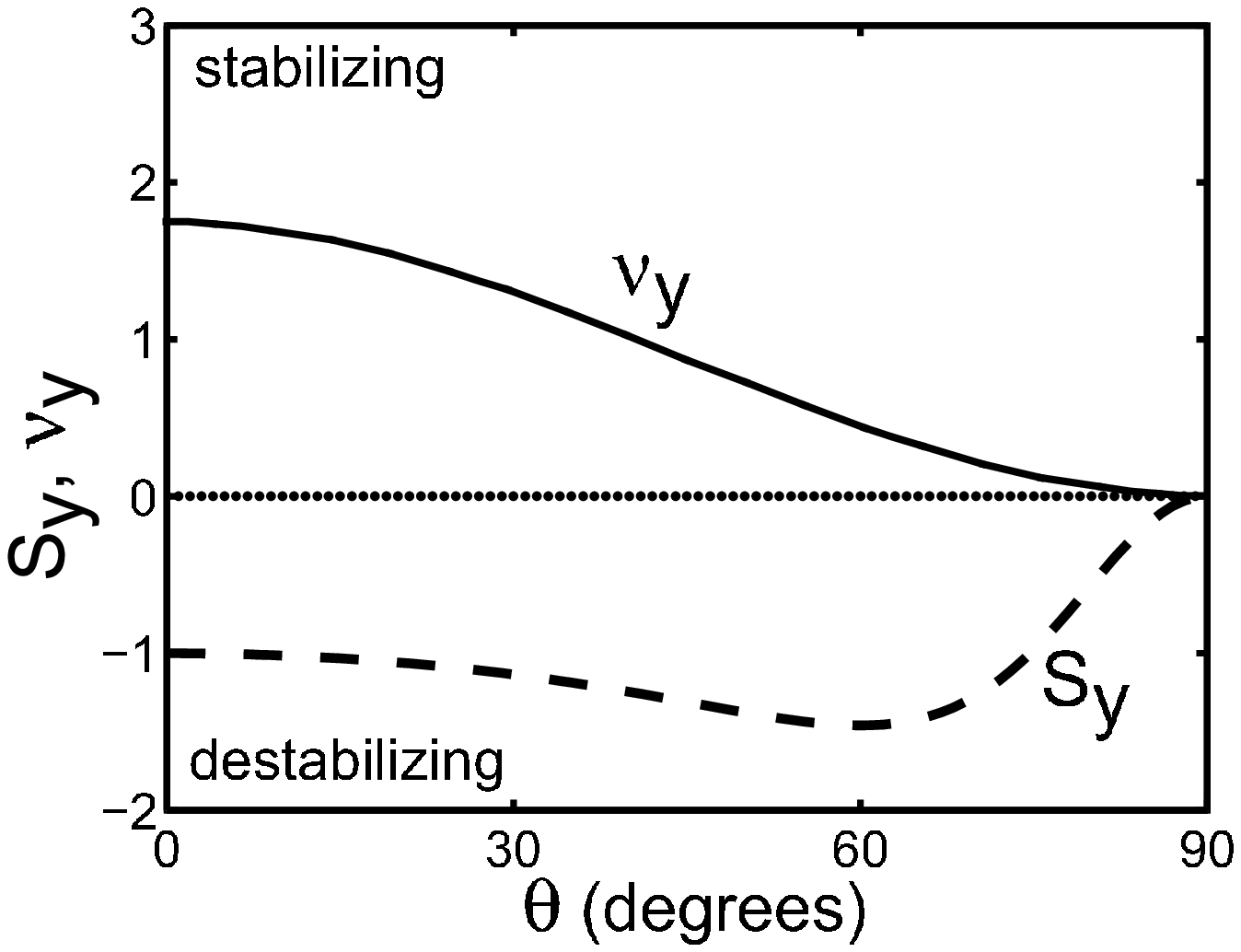}
\caption{\label{figCarter} Normalized coefficients $S_x,\nu_x,S_y$,
and $\nu_y$, comparing the effect of surface-induced currents
($\nu$, Eq. \ref{Moseler-term}) to erosion from Gaussian ellipsoids
($S$, Eq. \ref{eq0a}). The relative magnitude of $\nu/S$ at normal
incidence varies with conditions and is chosen arbitrarily here for
illustrative purposes.}
\end{center}
\end{figure}
Notably, the mechanism described by Eq. (\ref{Moseler-term})
corresponds to a conserved surface current and thus does not have any
effect on the yield curve $I(b)$. The effect of induced surface
currents on the stability is evident in Fig. \ref{figCarter}. The
effect stabilizes both modes from normal incidence up to incidence
angles of $45^o$, \textrm{whereupon it becomes a destabilizing
influence on only the longitudinal mode. The magnitudes of } $\nu_x$
and $\nu_y$ must equal each other at normal incidence, but their
relationship to the magnitudes of $S_x$ and $S_y$ depends on the
relative strengths of the mechanisms.  If the induced surface current
mechanism is sufficiently strong, as illustrated in Fig.
\ref{figCarter}, then starting with normal incidence and going to
increasing angles, one should observe a regime of absolute stability;
the dominance of parallel modes; and the dominance of perpendicular
modes. For further insight, it is essential to estimate the strength
of the induced surface current and how it depends on materials and
ion beam parameters, e.g. by methods such as atomistic simulations.

\section{Experimental Signatures}
We have described several mechanisms by which surface dynamics of the
form (\ref{eq0a}) can account for regions of ion beam angle where a
flat surface can be stable or unstable. The mechanisms suggested in
the previous two sections provide some scenarios leading to
modifications of the Bradley-Harper coefficients in Eq. (\ref{eq0a})
and thereby causing stability of the bombarded surface at various
ranges of angles; there are also potentially other such mechanisms.

The critical question now is to determine which of the potential
physical effects is operating in experiments; the answer to this
question almost certainly depends on the material, the ion mass and
energy, etc. Beyond the linear stability analysis itself, this issue
is of central importance for developing a quantitative nonlinear
theory of pattern formation; it is well known \cite{Cross93} that
accurately identifying the linear dispersion relation is critical for
deriving a nonlinear theory which can predict the fully developed
pattern.

How can experiments discern the dominant linear (in)stability
mechanism? Here we present one method for ruling out
some of the possibilities: in particular we point out the relevance of the
stability-instability transition not only as an interesting
dynamical phenomenon, but as a conceptual tool to gain valuable
information on the general character of the dynamics of
ion sputtered surfaces further away from the transition.

 In general, the linear stability
analyses discussed in this paper result in a dispersion relation of the form:
\begin{equation}
R_q \equiv Re \  (\omega_{\bf q}) =  - {S_x^{eff}} q_x^2 - {S_y^{eff}} q_y^2 -
B_{xx} q_x^4 - B_{yy} q_y^4  - B_{xy} q_x^2 q_y^2 + \cdots \ ,
\label{generalizedequation}
\end{equation}
\footnote{We assume the 4{\emph th} derivative term may acquire
anisotropic components, changing from $B\nabla^4 h$ to $B_{xx}
\partial^4 h /\partial x^4 + B_{yy} \partial^4 h /\partial y^4 +B_{xy}
\partial^4 h /\partial x^2 \partial y^2$. Anisotropy of the
fourth-order derivative term may emerge already within BH analysis,
as was noted by Makeev {\emph et al.}.}\cite{Makeev02} which
describes the growth rate of a Fourier mode:
\begin{equation}
\hat{h}_{q_x,q_y}(t) = \hat{h}_{q_x,q_y}(0) e^{i (q_x x + q_y y) +
\omega_{\bf q} t} \ . \label{fourierdynamics}
\end{equation}
In equation (\ref{generalizedequation}) we have lumped the two
quadratic contributions into $S_{x,y}^{eff}=S_{x,y}-\nu_{x,y}$. We
focus on the transition between stable and unstable perpendicular
(parallel) modes described by Eq. (\ref{generalizedequation}) as
$S_y^{eff}$ ($S_x^{eff} $) changes sign. This is depicted in the
left column of Fig. (4).
\begin{figure}
\begin{center}
\includegraphics[width=5.0in,clip=]{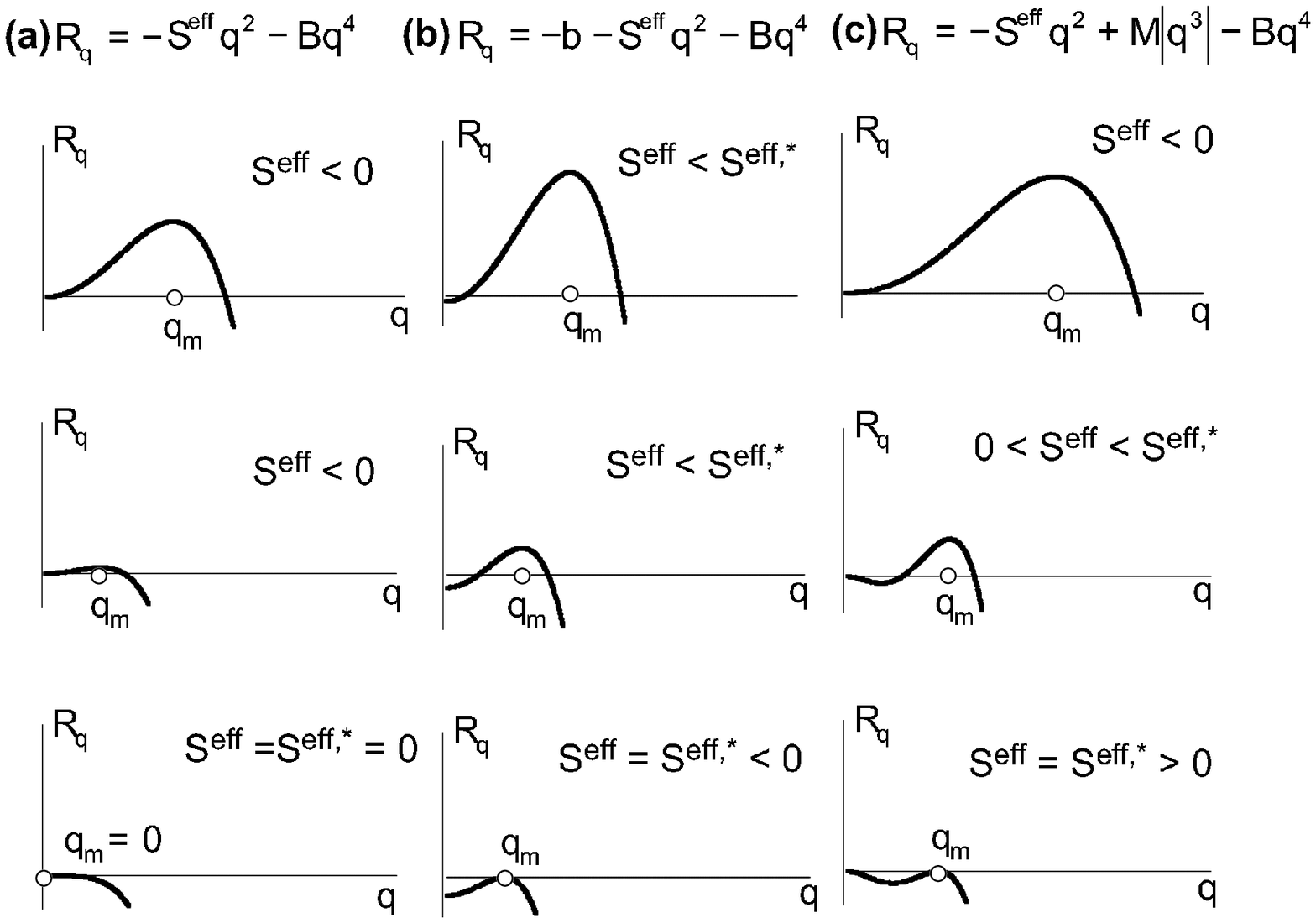}
\caption{\label{figshape} Schematic plots depicting the transition
between stable and unstable surface dynamics for three dispersion
relations. (a) Left column: generalized Bradley-Harper, Eq.
(\ref{generalizedequation}), where the transition occurs at
$S^{\rm{eff,*}} = 0$ with diverging wavelength. (b) Middle column:
with Facsko nonlocal ``damping term'', transition occurs at
$S^{\rm{eff,*}} < 0 $ with finite wavelength. (c) Right column: with
Asaro-Tiller nonlocal elastic energy mechanism, transition occurs at
$S^{\rm{eff,*}} > 0 $ with finite wavelength.}
\end{center}
\end{figure}
Here we assume for simplicity that the only parameters in Eq.
(\ref{generalizedequation}) that change appreciably with the beam
angle are $S_x^{eff},S_y^{eff}$.

The most important feature of this schematic
plot is that it predicts divergence of the pattern wavelength upon
reaching the transition to stable surface dynamics. To see this more
clearly, notice that a condition for the existence of linearly
unstable modes is that $\max(R_q)$, the maximal value of $R_q$ over
all wave vectors ${\bf q} = (q_x,q_y)$ is positive. Assuming a
smooth dependence of all coefficients on the beam angle, a
transition between stable and unstable surface dynamics corresponds
to a beam angle for which $\max (R_q) = 0$. For simplicity, let us
assume that $\max(R_q)$ is achieved for ${\bf q} = (q_{max},0)$.
Then: $q_{max} = \sqrt{-S_x^{eff}/2B_{xx}}$ and $\max(R_q) =
R_{(q_{max},0)} = - S_x^{eff}/4B_{xx} = 0$, implying $S_x^{eff} (\theta) \to
S^{*} = 0$ and hence $q_{max} \to 0$ at the transition.

A diverging length scale is a strong characteristic signature of the
stability-instability transition, and it is thus natural to ask
whether this prediction is valid if other physical processes, not
accounted for in this paper, influence the surface dynamics and thus
modify the dispersion relation (\ref{generalizedequation}). We argue
that this divergence is expected as long as the following
assumptions are satisfied:
\begin{enumerate}
\item The beam-angle dependence of all coefficients in the equation
is smooth.

\item The linear dynamics is analytic, ruling out terms like $|\nabla
h|$

\item The dynamics is first order in time.

\item Linear surface dynamics is local - namely, it can be described
by a partial differential equation (PDE).

\end{enumerate}
Assumption 1 is required because, as can be seen easily from Fig.
(4a), a discontinuous "jump" between negative and positive values of
$S_x^{eff},S_y^{eff}$ at some beam angle $\theta^*$ may yield a
transition to stable dynamics at $|q|>0$. Physically, a
discontinuous change of parameters is associated with abrupt changes
in material properties, such as amorphization of a crystalline
surface. For such a scenario to be associated with a smooth change
of the beam angle is sufficiently unlikely as to be a rare
occurrence. Assumption 2 is required in order to make a linear
stability analysis meaningful. If this assumption is violated then
the early stage surface dynamics of an initially flat surface is not
described by the dynamics of independently evolving Fourier modes
(\ref{fourierdynamics}). Assumption 3 is expected to hold as long as
inertia is neglected. Assumptions 3 and 4 together imply that the
the amplification rate $R_q$, which is the real part of the complex
eigen-frequency $\omega_q$, contains only even positive powers of
$q$. Namely, local processes, by which a change of surface height is
related to the variation of erosion or flux rates between a surface
point and its nearest neighbors can be described by spatial
derivatives of the function $h(x,y,t)$. In a dynamics that is first
order in time, the eigen-frequency $\omega_q$ in Eq.
(\ref{fourierdynamics}) thus equals a polynomial in $q$, where all
spatial derivatives with odd order (i.e. $\nabla h$, $\nabla^3 h$)
have imaginary coefficients, and thus do not contribute to the
amplification rate $R_q =  Re \ (\omega_q)$. Notice also that the
locality assumption rules out the existence of a constant term (i.e.
$\propto q^0$) in (\ref{fourierdynamics}). This is a consequence of
the invariance $h \to h + const.$. Therefore, a term $\propto
h(x,y,t)$ (i.e. without spatial derivatives) can appear in the
surface dynamics only as a combination respecting this invariance
such as $h(x,y,t) - \bar h(t)$, where $\bar h(t) = \int d{\bf x}
h({\bf x,t})$, and thus must be associated with some nonlocal
processes.

Thus, under these general assumptions (and neglecting the
possibility that spatial derivatives of order 6 or higher are
dominant in the dynamics), the amplification rate $R_q$ satisfies
Eq. (\ref{generalizedequation}), the stability-instability
transition is depicted by Fig. (4a), and the characteristic
wavelength at the transition is predicted to diverge.

Recently, Ziberi \cite{ZiberiThesis} and George \cite{GeorgeThesis}
have measured the pattern wavelength at several values of beam angles
near the transition to the stable region in silicon irradiated by
noble gas ions at temperatures where the surface should be amorphous
and isotropic. The measurements  indicate that the wavelength at the
transition remains finite, and may thus be a strong indication that
one of the above assumptions is violated. Anticipating that
assumptions 1-3 are still valid, we will discuss here two nonlocal
terms, whose introduction may render the wavelength at the transition
finite.

\subsection{Facsko ``damping'' term}
First, let us consider the effect of including a linear term, $\bar
K[h(x,y,t) - \bar h(t)]$ with $\bar h(t) = \int d{\bf x} h({\bf
x,t})$, in the surface dynamics, Eq. ({\ref{eq0a}). Such a term was
recently introduced by Facsko {\it et. al.} \cite{Facsko04} as a
possible way to obtain long range ordered patterns observed in the
fully nonlinear regime. The term is suggested to be a placeholder
for a model of redeposition. With such a term, a constant $\bar K$
is added to the right hand side of the dispersion relation
(\ref{generalizedequation}). This is consistent with the dispersion
relation measured by Brown and Erlebacher \cite{Brown05a} on Si(111)
at temperatures where it should remain crystalline, with singular
surface energetics (making the validity of assumption (2)
questionable). The effect of this term on the transition between
stable and unstable dynamics is depicted in the middle column of
Fig. (4), where it is demonstrated that the characteristic
wavelength does not diverge at the transition, as can be obtained
from the following analysis: Again, for simplicity we assume that
$\max(R_q)$ is achieved for ${\bf q} = (q_{max},0)$. Here again
$q_{max} = \sqrt{-S_x^{eff}/2B_{xx}}$ but $\max(R_q) =
R_{(q_{max},0)} = K - S_x^{eff}/4B_{xx} = 0$, implying $S_x^{eff}
(\theta) = S^{*} < 0 $ and hence $|q_{max}| >0 $ at the transition.
\subsection{Asaro-Tiller mechanism}
The Asaro-Tiller elastic energy driven mechanism \cite{Asaro72,
Grinfeld86} gives rise to instability of solid surfaces under
biaxial in-plane stress. Biaxial compressive stresses are known to
develop in the bombarded
solid\cite{Brongersma00,Hedler05,VanDillen05,Otani06,Kim06,Meek71,VanDillen03},
and this effect could be important in the surface dynamics. Assuming
a sinusoidal modulation of a free surface under biaxial compressive
stress, the tangential stress increases at the troughs (compression)
and decreases at the peaks (dilation) by an amount proportional to
the wavenumber of the modulation and to the applied stress
$\sigma_0$ in the solid. This increases the chemical potential at
the troughs compared to the peaks and drives a surface current from
the troughs to the peaks that further amplifies the modulation, thus
leading to instability. Including this effect in the surface
dynamics gives rise to a term $\propto M |q|^3$ on the RHS of Eq.
(\ref{generalizedequation}) \cite{Panat05}, where $M\propto
\sigma_{0}^{2}$. This term does not stem from local effects but
rather from nonlocal effects associated with reducing elastic energy
throughout the whole solid. The effect of such a term on the
transition from stable to unstable surface dynamics is depicted in
the right-hand column of Fig. 4. As usual, we simplify the analysis
by assuming that $\max(R_q)$ is achieved for ${\bf q} = (q_{max},0)$
and solve the two equations: (i) $R_q = 0$ and (ii) $\partial R_q /
\partial q= 0 $, from which we get $S_x^{eff} (\theta) = S^{*} = M^2
/ 4B_{xx} >0 $ and $|q_{max}| = M/2B_{xx}
> 0 $ at the transition.

In this analysis we have implicitly assumed that the transition from
stable to unstable dynamics is ``supercritical" - namely, that it is
triggered by infinitesimal perturbations, and thus associated with a
change of sign of $\max (R_q)$. It is also possible that the
transition is ``subcritical", and occurs at parameters for which the
linear stability analysis, Eq. (\ref{generalizedequation}) yields
$\max (R_q) < 0$. If the transition is subcritical, then the
characteristic wavelength may not diverge even if the linear
dispersion is of the form (\ref{generalizedequation}). It is
possible to discern supercritical from subcritical transitions by
probing signatures of hysteretic behavior (associated with
subcritical but not with supercritical transitions), and by
carefully analyzing the kinetics of pattern formation. A necessary
condition for the existence of a subcritical transition is that the
leading nonlinear contributions to the dynamics have a destabilizing
effect (unlike the stabilizing nonlinear terms derived in
\cite{Makeev02}). Because our analysis is restricted to the linear
dynamics we will not pursue this possibility further here.

\section{Conclusions}
While the possibility of producing patterned surfaces has attracted
significant attention recently, few experiments have focused on
regions in parameter space where dynamically stable, smooth surfaces
are observed.  The existence of these stable regions contradicts the
Bradley-Harper stability analysis, but this is only part of the
reason for their importance: we have argued in this paper that the
emergence of stable surfaces provides important insights into the
surface dynamics, that are critical for the development of a
nonlinear theory of pattern formation in any parameters regime of
ion sputtering. Our major messages are:
\begin{enumerate}
\item The Bradley-Harper prediction regarding the instability of
ion-bombarded surfaces to perpendicular mode ripples follows from a
broad class of purely erosive response functions. This robustness
may explain why the Bradley-Harper picture seems to describe
correctly many observations of pattern evolution on ion sputtered
surfaces.

\item Various types of non-erosive
response can change the sign of the coefficient of the second
spatial derivative and thereby change the stability of surfaces to
the emergence of large scale patterns. In particular, modifications
of the response can lead to linear stability of smooth surfaces at
various ranges of beam angles. These changes can be accompanied by
no observable modification of the yield curve. Evidence for such
modifications should thus come from atomistic simulations or from
experiments that are capable of probing the local surface response
to a single ion impact.

\item Careful analysis of qualitative features of the pattern near
the transition between stability and instability of a flat surface,
in particular the existence or lack of divergence of the pattern
wavelength at the transition, enable us to determine conclusively
whether nonlocal mechanisms significantly affect the surface
dynamics. The outcome of this analysis is extremely important:
because the existence of nonlocal terms qualitatively changes the
linear dispersion relation, they must be included in the surface
dynamics, even away from the transition regime.
\end{enumerate}

This paper focused on the linear dynamics of ion sputtered surfaces.
In order to predict and control the fully developed patterns it is
necessary to extend this to a nonlinear analysis. The existence of a
stable-unstable transition at a critical beam angle $\theta_c$
presents an excellent opportunity for quantitative predictions about
pattern formation. Typically, near such a transition only a few
Fourier modes are unstable, and the morphology of evolving patterns
can generally be described by a weakly nonlinear ``amplitude
equation", whose form is universal and is determined almost solely
by symmetry considerations \cite{Cross93}. In other contexts, such
amplitude equations have been enormously successful at predicting
the shape of the selected patterns and many more features of their
dynamics. Such an approach has not been tried so far for ion
sputtered surfaces, apparently because it has been assumed that
there is no continuous control parameter whose variation may change
the stability of flat surfaces. Recognizing that the beam angle is
exactly such a parameter, at least for certain surfaces and ion
types and energies, may enable the application of this invaluable
theoretical tool to quantitative study of pattern formation on ion
sputtered surfaces.

We hope that the theoretical directions outlined in this paper will
trigger experimental and computational work that will lead to better
understanding of the surface response to ion impact and its
relevance to large scale surface dynamics, and to better
characterization of the transition from stability to instability of
flat surfaces. We believe that such insights will be important to
the development of a quantitative theory that will predict whether
and what types of patterns are formed on a sputtered surface for a
given set of material and ion beam parameters.

{\bf Acknowledgements} The research of BD was supported through the
Harvard MRSEC; the research of MJA was supported through
DE-FG02-06ER46335, and the research of MPB was supported through
NSF-0605031. BD acknowledges NWO for a travel grant and the
hospitality of the Lorentz Institute at Leiden University, where
part of this work was done. We thank W. Van Saarloos for pointing
our attention to the importance of careful analysis of the
stable-unstable transition. We thank H. Bola George for helpful
discussions and for preparing the plots in Fig. 4. We thank N.
Kalyanasundaram, H.T. Johnson and J.B. Freund for helpful
discussions.


\begin{thebibliography}{56}
\expandafter\ifx\csname natexlab\endcsname\relax\def\natexlab#1{#1}\fi
\expandafter\ifx\csname bibnamefont\endcsname\relax
  \def\bibnamefont#1{#1}\fi
\expandafter\ifx\csname bibfnamefont\endcsname\relax
  \def\bibfnamefont#1{#1}\fi
\expandafter\ifx\csname citenamefont\endcsname\relax
  \def\citenamefont#1{#1}\fi
\expandafter\ifx\csname url\endcsname\relax
  \def\url#1{\texttt{#1}}\fi
\expandafter\ifx\csname urlprefix\endcsname\relax\def\urlprefix{URL }\fi
\providecommand{\bibinfo}[2]{#2}
\providecommand{\eprint}[2][]{\url{#2}}

\bibitem[{\citenamefont{Navez et~al.}(1962)\citenamefont{Navez, Chaperot, and
  Sella}}]{Navez62}
\bibinfo{author}{\bibfnamefont{M.}~\bibnamefont{Navez}},
  \bibinfo{author}{\bibfnamefont{D.}~\bibnamefont{Chaperot}}, \bibnamefont{and}
  \bibinfo{author}{\bibfnamefont{C.}~\bibnamefont{Sella}},
  \bibinfo{journal}{Comptes Rendus Hebdomadaires Des Seances De L Academie Des
  Sciences} \textbf{\bibinfo{volume}{254}}, \bibinfo{pages}{240}
  (\bibinfo{year}{1962}).

\bibitem[{\citenamefont{Carter et~al.}(1977)\citenamefont{Carter, Nobes, Paton,
  Williams, and Whitton}}]{Carter77}
\bibinfo{author}{\bibfnamefont{G.}~\bibnamefont{Carter}},
  \bibinfo{author}{\bibfnamefont{M.}~\bibnamefont{Nobes}},
  \bibinfo{author}{\bibfnamefont{F.}~\bibnamefont{Paton}},
  \bibinfo{author}{\bibfnamefont{J.}~\bibnamefont{Williams}}, \bibnamefont{and}
  \bibinfo{author}{\bibfnamefont{J.}~\bibnamefont{Whitton}},
  \bibinfo{journal}{Rad. Eff.} \textbf{\bibinfo{volume}{33}},
  \bibinfo{pages}{65} (\bibinfo{year}{1977}).

\bibitem[{\citenamefont{Lewis et~al.}(1980)\citenamefont{Lewis, Nobes, Carter,
  and Whitton}}]{Lewis80}
\bibinfo{author}{\bibfnamefont{G.}~\bibnamefont{Lewis}},
  \bibinfo{author}{\bibfnamefont{M.}~\bibnamefont{Nobes}},
  \bibinfo{author}{\bibfnamefont{G.}~\bibnamefont{Carter}}, \bibnamefont{and}
  \bibinfo{author}{\bibfnamefont{J.}~\bibnamefont{Whitton}},
  \bibinfo{journal}{Nuc. Inst. and Meth.} \textbf{\bibinfo{volume}{170}}
  (\bibinfo{year}{1980}).

\bibitem[{\citenamefont{Bradley and Harper}(1988)}]{Bradley88}
\bibinfo{author}{\bibfnamefont{R.~M.} \bibnamefont{Bradley}} \bibnamefont{and}
  \bibinfo{author}{\bibfnamefont{J.~M.~E.} \bibnamefont{Harper}},
  \bibinfo{journal}{J. Vac. Sci. Technol. A} \textbf{\bibinfo{volume}{6}},
  \bibinfo{pages}{2390} (\bibinfo{year}{1988}).

\bibitem[{\citenamefont{Malherbe}(1994)}]{Malherbe94}
\bibinfo{author}{\bibfnamefont{J.}~\bibnamefont{Malherbe}},
  \bibinfo{journal}{Critical Reviews in Solid State and Materials Sciences}
  \textbf{\bibinfo{volume}{19}}, \bibinfo{pages}{129} (\bibinfo{year}{1994}).

\bibitem[{\citenamefont{Mayer et~al.}(1994)\citenamefont{Mayer, Chason, and
  Howard}}]{Mayer94}
\bibinfo{author}{\bibfnamefont{M.}~\bibnamefont{Mayer}},
  \bibinfo{author}{\bibfnamefont{E.}~\bibnamefont{Chason}}, \bibnamefont{and}
  \bibinfo{author}{\bibfnamefont{A.}~\bibnamefont{Howard}},
  \bibinfo{journal}{J. Appl. Phys.} \textbf{\bibinfo{volume}{76}},
  \bibinfo{pages}{1633} (\bibinfo{year}{1994}).

\bibitem[{\citenamefont{Chason et~al.}(1994)\citenamefont{Chason, Mayer,
  Kellerman, McIlroy, and Howard}}]{Chason94a}
\bibinfo{author}{\bibfnamefont{E.}~\bibnamefont{Chason}},
  \bibinfo{author}{\bibfnamefont{T.}~\bibnamefont{Mayer}},
  \bibinfo{author}{\bibfnamefont{B.}~\bibnamefont{Kellerman}},
  \bibinfo{author}{\bibfnamefont{D.}~\bibnamefont{McIlroy}}, \bibnamefont{and}
  \bibinfo{author}{\bibfnamefont{A.}~\bibnamefont{Howard}},
  \bibinfo{journal}{Phys. Rev. Lett.} \textbf{\bibinfo{volume}{72}},
  \bibinfo{pages}{3040} (\bibinfo{year}{1994}).

\bibitem[{\citenamefont{Cuerno and Barabasi}(1995)}]{Cuerno95}
\bibinfo{author}{\bibfnamefont{R.}~\bibnamefont{Cuerno}} \bibnamefont{and}
  \bibinfo{author}{\bibfnamefont{A.-L.} \bibnamefont{Barabasi}},
  \bibinfo{journal}{Phys. Rev. Lett.} \textbf{\bibinfo{volume}{74}},
  \bibinfo{pages}{4746 } (\bibinfo{year}{1995}).

\bibitem[{\citenamefont{Carter and Vishnyakov}(1996)}]{Carter96}
\bibinfo{author}{\bibfnamefont{G.}~\bibnamefont{Carter}} \bibnamefont{and}
  \bibinfo{author}{\bibfnamefont{V.}~\bibnamefont{Vishnyakov}},
  \bibinfo{journal}{Phys. Rev. B} \textbf{\bibinfo{volume}{54}},
  \bibinfo{pages}{17647} (\bibinfo{year}{1996}).

\bibitem[{\citenamefont{Makeev and Barabasi}(1997)}]{Makeev97}
\bibinfo{author}{\bibfnamefont{M.}~\bibnamefont{Makeev}} \bibnamefont{and}
  \bibinfo{author}{\bibfnamefont{A.-L.} \bibnamefont{Barabasi}},
  \bibinfo{journal}{Appl. Phys. Lett.} \textbf{\bibinfo{volume}{71}},
  \bibinfo{pages}{2800} (\bibinfo{year}{1997}).

\bibitem[{\citenamefont{Facsko et~al.}(1999)\citenamefont{Facsko, Dekorsy,
  Koerdt, Trappe, Kurz, Vogt, and Hartnagel}}]{Facsko99}
\bibinfo{author}{\bibfnamefont{S.}~\bibnamefont{Facsko}},
  \bibinfo{author}{\bibfnamefont{T.}~\bibnamefont{Dekorsy}},
  \bibinfo{author}{\bibfnamefont{C.}~\bibnamefont{Koerdt}},
  \bibinfo{author}{\bibfnamefont{C.}~\bibnamefont{Trappe}},
  \bibinfo{author}{\bibfnamefont{H.}~\bibnamefont{Kurz}},
  \bibinfo{author}{\bibfnamefont{A.}~\bibnamefont{Vogt}}, \bibnamefont{and}
  \bibinfo{author}{\bibfnamefont{H.}~\bibnamefont{Hartnagel}},
  \bibinfo{journal}{Science} \textbf{\bibinfo{volume}{285}},
  \bibinfo{pages}{1551} (\bibinfo{year}{1999}).

\bibitem[{\citenamefont{Judy et~al.}(1999)\citenamefont{Judy, Murty, Butler,
  Pomeroy, Cooper, Woll, Brock, Kycia, and Headrick}}]{Judy99}
\bibinfo{author}{\bibfnamefont{A.}~\bibnamefont{Judy}},
  \bibinfo{author}{\bibfnamefont{M.~R.} \bibnamefont{Murty}},
  \bibinfo{author}{\bibfnamefont{E.}~\bibnamefont{Butler}},
  \bibinfo{author}{\bibfnamefont{J.}~\bibnamefont{Pomeroy}},
  \bibinfo{author}{\bibfnamefont{B.}~\bibnamefont{Cooper}},
  \bibinfo{author}{\bibfnamefont{A.}~\bibnamefont{Woll}},
  \bibinfo{author}{\bibfnamefont{J.}~\bibnamefont{Brock}},
  \bibinfo{author}{\bibfnamefont{S.}~\bibnamefont{Kycia}}, \bibnamefont{and}
  \bibinfo{author}{\bibfnamefont{R.}~\bibnamefont{Headrick}},
  \bibinfo{journal}{Mater. Res. Soc. Symp. Proc.}
  \textbf{\bibinfo{volume}{570}}, \bibinfo{pages}{61} (\bibinfo{year}{1999}).

\bibitem[{\citenamefont{Erlebacher et~al.}(1999)\citenamefont{Erlebacher, Aziz,
  Chason, Sinclair, and Floro}}]{Erlebacher99}
\bibinfo{author}{\bibfnamefont{J.}~\bibnamefont{Erlebacher}},
  \bibinfo{author}{\bibfnamefont{M.~J.} \bibnamefont{Aziz}},
  \bibinfo{author}{\bibfnamefont{E.}~\bibnamefont{Chason}},
  \bibinfo{author}{\bibfnamefont{M.~B.} \bibnamefont{Sinclair}},
  \bibnamefont{and} \bibinfo{author}{\bibfnamefont{J.}~\bibnamefont{Floro}},
  \bibinfo{journal}{Phys. Rev. Lett.} \textbf{\bibinfo{volume}{82}},
  \bibinfo{pages}{2330} (\bibinfo{year}{1999}).

\bibitem[{\citenamefont{Erlebacher et~al.}(2000)\citenamefont{Erlebacher, Aziz,
  Chason, Sinclair, and Floro}}]{Erlebacher00}
\bibinfo{author}{\bibfnamefont{J.}~\bibnamefont{Erlebacher}},
  \bibinfo{author}{\bibfnamefont{M.~J.} \bibnamefont{Aziz}},
  \bibinfo{author}{\bibfnamefont{E.}~\bibnamefont{Chason}},
  \bibinfo{author}{\bibfnamefont{M.~B.} \bibnamefont{Sinclair}},
  \bibnamefont{and} \bibinfo{author}{\bibfnamefont{J.~A.} \bibnamefont{Floro}},
  \bibinfo{journal}{J. Vac. Sci. Technol. A} \textbf{\bibinfo{volume}{18}},
  \bibinfo{pages}{115} (\bibinfo{year}{2000}).

\bibitem[{\citenamefont{Facsko et~al.}(2001)\citenamefont{Facsko, Bobek,
  Dekorsy, and Kurz}}]{Facsko01}
\bibinfo{author}{\bibfnamefont{S.}~\bibnamefont{Facsko}},
  \bibinfo{author}{\bibfnamefont{T.}~\bibnamefont{Bobek}},
  \bibinfo{author}{\bibfnamefont{T.}~\bibnamefont{Dekorsy}}, \bibnamefont{and}
  \bibinfo{author}{\bibfnamefont{H.}~\bibnamefont{Kurz}},
  \bibinfo{journal}{Phys. Stat. Sol. B} \textbf{\bibinfo{volume}{224}},
  \bibinfo{pages}{537} (\bibinfo{year}{2001}).

\bibitem[{\citenamefont{Flamm et~al.}(2001)\citenamefont{Flamm, Frost, and
  Hirsch}}]{Flamm01}
\bibinfo{author}{\bibfnamefont{D.}~\bibnamefont{Flamm}},
  \bibinfo{author}{\bibfnamefont{F.}~\bibnamefont{Frost}}, \bibnamefont{and}
  \bibinfo{author}{\bibfnamefont{D.}~\bibnamefont{Hirsch}},
  \bibinfo{journal}{Appl. Surf. Sci.} \textbf{\bibinfo{volume}{179}},
  \bibinfo{pages}{95} (\bibinfo{year}{2001}).

\bibitem[{\citenamefont{Habenicht}(2001)}]{Habenicht01}
\bibinfo{author}{\bibfnamefont{S.}~\bibnamefont{Habenicht}},
  \bibinfo{journal}{Phys. Rev. B} \textbf{\bibinfo{volume}{63}},
  \bibinfo{pages}{125419} (\bibinfo{year}{2001}).

\bibitem[{\citenamefont{Costantini et~al.}(2001)\citenamefont{Costantini,
  de~Mongeot, Boragno, and Valbusa}}]{Costantini01}
\bibinfo{author}{\bibfnamefont{G.}~\bibnamefont{Costantini}},
  \bibinfo{author}{\bibfnamefont{F.~B.} \bibnamefont{de~Mongeot}},
  \bibinfo{author}{\bibfnamefont{C.}~\bibnamefont{Boragno}}, \bibnamefont{and}
  \bibinfo{author}{\bibfnamefont{U.}~\bibnamefont{Valbusa}},
  \bibinfo{journal}{Phys. Rev. Lett.} \textbf{\bibinfo{volume}{86}},
  \bibinfo{pages}{838} (\bibinfo{year}{2001}).

\bibitem[{\citenamefont{Umbach et~al.}(2001)\citenamefont{Umbach, Headrick, and
  Chang}}]{Umbach01}
\bibinfo{author}{\bibfnamefont{C.}~\bibnamefont{Umbach}},
  \bibinfo{author}{\bibfnamefont{R.}~\bibnamefont{Headrick}}, \bibnamefont{and}
  \bibinfo{author}{\bibfnamefont{K.}~\bibnamefont{Chang}},
  \bibinfo{journal}{Phys. Rev. Lett.} \textbf{\bibinfo{volume}{87}},
  \bibinfo{pages}{246104} (\bibinfo{year}{2001}).

\bibitem[{\citenamefont{Valbusa et~al.}(2002)\citenamefont{Valbusa, Boragno,
  and de~Mongeot}}]{Valbusa02}
\bibinfo{author}{\bibfnamefont{U.}~\bibnamefont{Valbusa}},
  \bibinfo{author}{\bibfnamefont{C.}~\bibnamefont{Boragno}}, \bibnamefont{and}
  \bibinfo{author}{\bibfnamefont{F.~B.} \bibnamefont{de~Mongeot}},
  \bibinfo{journal}{J. Phys.: Condens. Matter} \textbf{\bibinfo{volume}{14}},
  \bibinfo{pages}{8153} (\bibinfo{year}{2002}).

\bibitem[{\citenamefont{Makeev et~al.}(2002)\citenamefont{Makeev, Cuerno, and
  Barabasi}}]{Makeev02}
\bibinfo{author}{\bibfnamefont{M.~A.} \bibnamefont{Makeev}},
  \bibinfo{author}{\bibfnamefont{R.}~\bibnamefont{Cuerno}}, \bibnamefont{and}
  \bibinfo{author}{\bibfnamefont{A.-L.} \bibnamefont{Barabasi}},
  \bibinfo{journal}{Nucl. Instr. and Meth. in Phys. Res. B}
  \textbf{\bibinfo{volume}{197}}, \bibinfo{pages}{185} (\bibinfo{year}{2002}).

\bibitem[{\citenamefont{Chason and Aziz}(2003)}]{Chason03}
\bibinfo{author}{\bibfnamefont{E.}~\bibnamefont{Chason}} \bibnamefont{and}
  \bibinfo{author}{\bibfnamefont{M.J.}~\bibnamefont{Aziz}},
  \bibinfo{journal}{Scripta Mater.} \textbf{\bibinfo{volume}{49}},
  \bibinfo{pages}{953} (\bibinfo{year}{2003}).

\bibitem[{\citenamefont{Facsko et~al.}(2004)\citenamefont{Facsko, Bobek, Kurz,
  and Dekorsy}}]{Facsko04}
\bibinfo{author}{\bibfnamefont{S.}~\bibnamefont{Facsko}},
  \bibinfo{author}{\bibfnamefont{T.}~\bibnamefont{Bobek}},
  \bibinfo{author}{\bibfnamefont{A. Stahl}, \bibnamefont{H. Kurz}},
  \bibnamefont{and} \bibinfo{author}{\bibfnamefont{T.}~\bibnamefont{Dekorsy}},
  \bibinfo{journal}{Physical Review B} \textbf{\bibinfo{volume}{69}}
  (\bibinfo{year}{2004}).

\bibitem[{\citenamefont{Castro et~al.}(2005)\citenamefont{Castro, Cuerno,
  Vazquez, and Gago}}]{Castro05}
\bibinfo{author}{\bibfnamefont{M.}~\bibnamefont{Castro}},
  \bibinfo{author}{\bibfnamefont{R.}~\bibnamefont{Cuerno}},
  \bibinfo{author}{\bibfnamefont{L.}~\bibnamefont{Vazquez}}, \bibnamefont{and}
  \bibinfo{author}{\bibfnamefont{R.}~\bibnamefont{Gago}},
  \bibinfo{journal}{Phys. Rev. Lett.} \textbf{\bibinfo{volume}{94}},
  \bibinfo{pages}{016102} (\bibinfo{year}{2005}).

\bibitem[{\citenamefont{Chan and Chason}(2005)}]{Chan05}
\bibinfo{author}{\bibfnamefont{W.}~\bibnamefont{Chan}} \bibnamefont{and}
  \bibinfo{author}{\bibfnamefont{E.}~\bibnamefont{Chason}},
  \bibinfo{journal}{Phys. Rev. B} \textbf{\bibinfo{volume}{72}},
  \bibinfo{pages}{165418} (\bibinfo{year}{2005}).

\bibitem[{\citenamefont{Cuenat et~al.}(2005)\citenamefont{Cuenat, George,
  Chang, Blakely, and Aziz}}]{Cuenat05}
\bibinfo{author}{\bibfnamefont{A.}~\bibnamefont{Cuenat}},
  \bibinfo{author}{\bibfnamefont{H.~B.} \bibnamefont{George}},
  \bibinfo{author}{\bibfnamefont{K.-C.} \bibnamefont{Chang}},
  \bibinfo{author}{\bibfnamefont{J.}~\bibnamefont{Blakely}}, \bibnamefont{and}
  \bibinfo{author}{\bibfnamefont{M.~J.} \bibnamefont{Aziz}},
  \bibinfo{journal}{Adv. Mater.} \textbf{\bibinfo{volume}{17}},
  \bibinfo{pages}{2845} (\bibinfo{year}{2005}).

\bibitem[{\citenamefont{Feix et~al.}(2005)\citenamefont{Feix, Hartmann, Kree,
  Munoz-Garcia, and Cuerno}}]{Feix05}
\bibinfo{author}{\bibfnamefont{M.}~\bibnamefont{Feix}},
  \bibinfo{author}{\bibfnamefont{A.}~\bibnamefont{Hartmann}},
  \bibinfo{author}{\bibfnamefont{R.}~\bibnamefont{Kree}},
  \bibinfo{author}{\bibfnamefont{J.}~\bibnamefont{Munoz-Garcia}},
  \bibnamefont{and} \bibinfo{author}{\bibfnamefont{R.}~\bibnamefont{Cuerno}},
  \bibinfo{journal}{Phys. Rev. B} \textbf{\bibinfo{volume}{71}}
  (\bibinfo{year}{2005}).

\bibitem[{\citenamefont{Rusponi et~al.}(1997)\citenamefont{Rusponi, Boragno,
  and Valbusa}}]{Rusponi05}
\bibinfo{author}{\bibfnamefont{S.}~\bibnamefont{Rusponi}},
  \bibinfo{author}{\bibfnamefont{C.}~\bibnamefont{Boragno}}, \bibnamefont{and}
  \bibinfo{author}{\bibfnamefont{U.}~\bibnamefont{Valbusa}},
  \bibinfo{journal}{Phys. Rev. Lett.} \textbf{\bibinfo{volume}{78}}
  (\bibinfo{year}{1997}).

\bibitem[{\citenamefont{Brown and Erlebacher}(2005)}]{Brown05a}
\bibinfo{author}{\bibfnamefont{A.}~\bibnamefont{Brown}} \bibnamefont{and}
  \bibinfo{author}{\bibfnamefont{J.}~\bibnamefont{Erlebacher}},
  \bibinfo{journal}{Phys. Rev. B} \textbf{\bibinfo{volume}{72}},
  \bibinfo{pages}{075350} (\bibinfo{year}{2005}).

\bibitem[{\citenamefont{Brown et~al.}(2005)\citenamefont{Brown, Erlebacher,
  Chan, and Chason}}]{Brown05b}
\bibinfo{author}{\bibfnamefont{A.}~\bibnamefont{Brown}},
  \bibinfo{author}{\bibfnamefont{J.}~\bibnamefont{Erlebacher}},
  \bibinfo{author}{\bibfnamefont{W.}~\bibnamefont{Chan}}, \bibnamefont{and}
  \bibinfo{author}{\bibfnamefont{E.}~\bibnamefont{Chason}},
  \bibinfo{journal}{Phys. Rev. Lett.} \textbf{\bibinfo{volume}{95}},
  \bibinfo{pages}{056101} (\bibinfo{year}{2005}).

\bibitem[{\citenamefont{Ziberi et~al.}(2005)\citenamefont{Ziberi, Frost, Hoche,
  and Rauschenbach}}]{Ziberi05}
\bibinfo{author}{\bibfnamefont{B.}~\bibnamefont{Ziberi}},
  \bibinfo{author}{\bibfnamefont{F.}~\bibnamefont{Frost}},
  \bibinfo{author}{\bibfnamefont{T.}~\bibnamefont{Hoche}}, \bibnamefont{and}
  \bibinfo{author}{\bibfnamefont{B.}~\bibnamefont{Rauschenbach}},
  \bibinfo{journal}{Phys. Rev. B} \textbf{\bibinfo{volume}{72}},
  \bibinfo{pages}{235310} (\bibinfo{year}{2005}).

\bibitem[{\citenamefont{Ziberi et~al.}(2006)\citenamefont{Ziberi, Frost, and
  Rauschenbach}}]{Ziberi06a}
\bibinfo{author}{\bibfnamefont{B.}~\bibnamefont{Ziberi}},
  \bibinfo{author}{\bibfnamefont{F.}~\bibnamefont{Frost}}, \bibnamefont{and}
  \bibinfo{author}{\bibfnamefont{B.}~\bibnamefont{Rauschenbach}},
  \bibinfo{journal}{J.Vac. Sci. Technol. A} \textbf{\bibinfo{volume}{24}},
  \bibinfo{pages}{1344} (\bibinfo{year}{2006}).

\bibitem[{\citenamefont{Aziz}(2006)}]{Aziz06}
\bibinfo{author}{\bibfnamefont{M. J.}~\bibnamefont{Aziz}}, \bibinfo{journal}{Mat.
  Fys. Medd. Dan. Vid. Selsk.} \textbf{\bibinfo{volume}{52}} \bibinfo{pages}{187}
  (\bibinfo{year}{2006}).

\bibitem[{\citenamefont{Sigmund}(1969)}]{Sigmund69}
\bibinfo{author}{\bibfnamefont{P.}~\bibnamefont{Sigmund}},
 \bibinfo{journal}{Phys. Rev.}  \textbf{\bibinfo{volume}{184}}, \bibinfo{pages}{383} (\bibinfo{year}{1969}).

\bibitem[{\citenamefont{Sigmund}(1973)}]{Sigmund73}
\bibinfo{author}{\bibfnamefont{P.}~\bibnamefont{Sigmund}}, \bibinfo{journal}{J.
  Mater. Sci.} \textbf{\bibinfo{volume}{8}}, \bibinfo{pages}{1545}
  (\bibinfo{year}{1973}).

\bibitem[{\citenamefont{Herring}(1950)}]{Herring50}
\bibinfo{author}{\bibfnamefont{C.}~\bibnamefont{Herring}}, \bibinfo{journal}{J.
  Appl. Phys.} \textbf{\bibinfo{volume}{21}}, \bibinfo{pages}{301}
  (\bibinfo{year}{1950}).

\bibitem[{\citenamefont{Mullins}(1959)}]{Mullins59}
\bibinfo{author}{\bibfnamefont{W.}~\bibnamefont{Mullins}}, \bibinfo{journal}{J.
  Appl. Phys.} \textbf{\bibinfo{volume}{30}}, \bibinfo{pages}{77}
  (\bibinfo{year}{1959}).

\bibitem[{\citenamefont{Cross and Hohenberg}(1993)}]{Cross93}
\bibinfo{author}{\bibfnamefont{M.}~\bibnamefont{Cross}} \bibnamefont{and}
  \bibinfo{author}{\bibfnamefont{P.}~\bibnamefont{Hohenberg}},
  \bibinfo{journal}{Rev. Mod. Phys.} \textbf{\bibinfo{volume}{65}}
  (\bibinfo{year}{1993}).

\bibitem[{\citenamefont{Yamada et~al.}(2001)\citenamefont{Yamada, Matsuo,
  Toyoda, and Kirkpatrick}}]{Yamada01}
\bibinfo{author}{\bibfnamefont{I.}~\bibnamefont{Yamada}},
  \bibinfo{author}{\bibfnamefont{J.}~\bibnamefont{Matsuo}},
  \bibinfo{author}{\bibfnamefont{N.}~\bibnamefont{Toyoda}}, \bibnamefont{and}
  \bibinfo{author}{\bibfnamefont{A.}~\bibnamefont{Kirkpatrick}},
  \bibinfo{journal}{Mater. Sci. Eng. R-Reports} \textbf{\bibinfo{volume}{34}},
  \bibinfo{pages}{231} (\bibinfo{year}{2001}).

\bibitem[{\citenamefont{Bringa et~al.}(2001)\citenamefont{Bringa, Nordlund, and
  Keinonen}}]{Bringa01}
\bibinfo{author}{\bibfnamefont{E.}~\bibnamefont{Bringa}},
  \bibinfo{author}{\bibfnamefont{K.}~\bibnamefont{Nordlund}}, \bibnamefont{and}
  \bibinfo{author}{\bibfnamefont{J.}~\bibnamefont{Keinonen}},
  \bibinfo{journal}{Phys. Rev. B} \textbf{\bibinfo{volume}{64}},
  \bibinfo{pages}{235426} (\bibinfo{year}{2001}).

\bibitem[{\citenamefont{Kalyanasundaram
  et~al.}(2007)\citenamefont{Kalyanasundaram, Johnson, and Freund}}]{Naga07}
\bibinfo{author}{\bibfnamefont{N.}~\bibnamefont{Kalyanasundaram}},
  \bibinfo{author}{\bibfnamefont{H.}~\bibnamefont{Johnson}}, \bibnamefont{and}
  \bibinfo{author}{\bibfnamefont{J.}~\bibnamefont{Freund}},
  \bibinfo{journal}{unpublished}.

\bibitem[{\citenamefont{Moseler et~al.}(2005)\citenamefont{Moseler, Gumbsch,
  Casiraghi, Ferrari, and Robertson}}]{Moseler05}
\bibinfo{author}{\bibfnamefont{M.}~\bibnamefont{Moseler}},
  \bibinfo{author}{\bibfnamefont{P.}~\bibnamefont{Gumbsch}},
  \bibinfo{author}{\bibfnamefont{C.}~\bibnamefont{Casiraghi}},
  \bibinfo{author}{\bibfnamefont{A.}~\bibnamefont{Ferrari}}, \bibnamefont{and}
  \bibinfo{author}{\bibfnamefont{J.}~\bibnamefont{Robertson}},
  \bibinfo{journal}{Science} \textbf{\bibinfo{volume}{309}},
  \bibinfo{pages}{1545} (\bibinfo{year}{2005}).

\bibitem[{\citenamefont{Vasile et~al.}(1999)\citenamefont{Vasile, Xie, and
  Nassar}}]{vasile99}
\bibinfo{author}{\bibfnamefont{M.}~\bibnamefont{Vasile}},
  \bibinfo{author}{\bibfnamefont{J.}~\bibnamefont{Xie}}, \bibnamefont{and}
  \bibinfo{author}{\bibfnamefont{R.}~\bibnamefont{Nassar}},  \bibinfo{journal}{J. Vac. Sci. Technol. B}
  \textbf{\bibinfo{volume}{17}}, \bibinfo{pages}{3085} (\bibinfo{year}{1999}).

\bibitem[{\citenamefont{Ziberi}(2006)}]{ZiberiThesis}
\bibinfo{author}{\bibfnamefont{B.}~\bibnamefont{Ziberi}}
\bibinfo{journal}{Ph.D. Thesis,  University of Leipzig}
  (\bibinfo{year}{2006}).

\bibitem[{\citenamefont{George}(2007)}]{GeorgeThesis}
\bibinfo{author}{\bibfnamefont{H.}~\bibnamefont{George}},
  \bibinfo{journal}{Ph.D. Thesis, Harvard University}  (\bibinfo{year}{2007}).

\bibitem[{\citenamefont{Asaro and Tiller}(1972)}]{Asaro72}
\bibinfo{author}{\bibfnamefont{R.}~\bibnamefont{Asaro}} \bibnamefont{and}
  \bibinfo{author}{\bibfnamefont{W.}~\bibnamefont{Tiller}},
  \bibinfo{journal}{Metall. Trans.} \textbf{\bibinfo{volume}{3}},
  \bibinfo{pages}{1789} (\bibinfo{year}{1972}).

\bibitem[{\citenamefont{Grinfeld}(1086)}]{Grinfeld86}
\bibinfo{author}{\bibfnamefont{M.}~\bibnamefont{Grinfeld}},
  \bibinfo{journal}{Sov. Phys. Dokl.} \textbf{\bibinfo{volume}{31}},
  \bibinfo{pages}{831} (\bibinfo{year}{1086}).

\bibitem[{\citenamefont{Brongersma et~al.}(2002)\citenamefont{Brongersma,
  Snoeks, van Dillen, and Polman}}]{Brongersma00}
\bibinfo{author}{\bibfnamefont{M.}~\bibnamefont{Brongersma}},
  \bibinfo{author}{\bibfnamefont{E.}~\bibnamefont{Snoeks}},
  \bibinfo{author}{\bibfnamefont{T.}~\bibnamefont{van Dillen}},
  \bibnamefont{and} \bibinfo{author}{\bibfnamefont{A.}~\bibnamefont{Polman}},
  \bibinfo{journal}{J. Appl. Phys.} \textbf{\bibinfo{volume}{88}}
  (\bibinfo{year}{2002}).

\bibitem[{\citenamefont{Hedler et~al.}(2005)\citenamefont{Hedler, Klaumunzer,
  and Wesch}}]{Hedler05}
\bibinfo{author}{\bibfnamefont{A.}~\bibnamefont{Hedler}},
  \bibinfo{author}{\bibfnamefont{S.}~\bibnamefont{Klaumunzer}},
  \bibnamefont{and} \bibinfo{author}{\bibfnamefont{W.}~\bibnamefont{Wesch}},
  \bibinfo{journal}{Phys. Rev. B} \textbf{\bibinfo{volume}{72}}
  (\bibinfo{year}{2005}).

\bibitem[{\citenamefont{van Dillen et~al.}(2005)\citenamefont{van Dillen,
  Polman, Onck, and van~der Giessen}}]{VanDillen05}
\bibinfo{author}{\bibfnamefont{T.}~\bibnamefont{van Dillen}},
  \bibinfo{author}{\bibfnamefont{A.}~\bibnamefont{Polman}},
  \bibinfo{author}{\bibfnamefont{P.}~\bibnamefont{Onck}}, \bibnamefont{and}
  \bibinfo{author}{\bibfnamefont{E.}~\bibnamefont{van~der Giessen}},
  \bibinfo{journal}{Phys. Rev. B} \textbf{\bibinfo{volume}{71}},
  \bibinfo{pages}{024103} (\bibinfo{year}{2005}).

\bibitem[{\citenamefont{Otani et~al.}(2006)\citenamefont{Otani, Chen,
  Hutchinson, Chervinsky, and Aziz}}]{Otani06}
\bibinfo{author}{\bibfnamefont{K.}~\bibnamefont{Otani}},
  \bibinfo{author}{\bibfnamefont{X.}~\bibnamefont{Chen}},
  \bibinfo{author}{\bibfnamefont{J.}~\bibnamefont{Hutchinson}},
  \bibinfo{author}{\bibfnamefont{J.}~\bibnamefont{Chervinsky}},
  \bibnamefont{and} \bibinfo{author}{\bibfnamefont{M.J.}~\bibnamefont{Aziz}},
  \bibinfo{journal}{J. Appl. Phys.} \textbf{\bibinfo{volume}{100}},
  \bibinfo{pages}{023535} (\bibinfo{year}{2006}).

\bibitem[{\citenamefont{Kim et~al.}(2006)\citenamefont{Kim, Chen, Aziz,
  Branton, and Vlassak}}]{Kim06}
\bibinfo{author}{\bibfnamefont{Y.-R.} \bibnamefont{Kim}},
  \bibinfo{author}{\bibfnamefont{P.}~\bibnamefont{Chen}},
  \bibinfo{author}{\bibfnamefont{M.J.}~\bibnamefont{Aziz}},
  \bibinfo{author}{\bibfnamefont{D.}~\bibnamefont{Branton}}, \bibnamefont{and}
  \bibinfo{author}{\bibfnamefont{J.J.}~\bibnamefont{Vlassak}},
  \bibinfo{journal}{J. Appl. Phys.}   \textbf{\bibinfo{volume}{100}} (\bibinfo{year}{2006}).

\bibitem[{\citenamefont{Meek et~al.}(1971)\citenamefont{Meek, Gibson, and
  Sellscho}}]{Meek71}
\bibinfo{author}{\bibfnamefont{R.}~\bibnamefont{Meek}},
  \bibinfo{author}{\bibfnamefont{W.}~\bibnamefont{Gibson}}, \bibnamefont{and}
  \bibinfo{author}{\bibfnamefont{J.}~\bibnamefont{Sellscho}},
  \bibinfo{journal}{Applied Physics Letters} \textbf{\bibinfo{volume}{18}},
  \bibinfo{pages}{535} (\bibinfo{year}{1971}).

\bibitem[{\citenamefont{van Dillen et~al.}(2003)\citenamefont{van Dillen,
  Polman, and van Kats}}]{VanDillen03}
\bibinfo{author}{\bibfnamefont{T.}~\bibnamefont{van Dillen}},
  \bibinfo{author}{\bibfnamefont{A.}~\bibnamefont{Polman}}, \bibnamefont{and}
  \bibinfo{author}{\bibfnamefont{C.~M.} \bibnamefont{van Kats}},
  \bibinfo{journal}{Applied Physics Letters} \textbf{\bibinfo{volume}{83}},
  \bibinfo{pages}{4315} (\bibinfo{year}{2003}).

\bibitem[{\citenamefont{Panat et~al.}(2005)\citenamefont{Panat, Hsia, and
  Cahill}}]{Panat05}
\bibinfo{author}{\bibfnamefont{R.}~\bibnamefont{Panat}},
  \bibinfo{author}{\bibfnamefont{K.}~\bibnamefont{Hsia}}, \bibnamefont{and}
  \bibinfo{author}{\bibfnamefont{D.}~\bibnamefont{Cahill}},
  \bibinfo{journal}{J. Appl. Phys.} \textbf{\bibinfo{volume}{97}}
  (\bibinfo{year}{2005}).

\bibitem[{\citenamefont{Villain}(1991)}]{Villain91}
\bibinfo{author}{\bibfnamefont{J.}~\bibnamefont{Villain}},
  \bibinfo{journal}{Journal De Physique I} \textbf{\bibinfo{volume}{1}},
  \bibinfo{pages}{19} (\bibinfo{year}{1991}).

\end{thebibliography}
\end{document}